\documentclass[twocolumn,aps,prb,amsmath,amsfonts,amssymb,floatfix,letterpaper,showpacs]{revtex4}

\pdfoutput=1

\usepackage{graphicx}
\usepackage{ifpdf}

\usepackage{hyperref}
\hypersetup{pdftitle={Higgs transitions of spin ice},pdfauthor={Stephen Powell}}

\newcommand{\ee}{\mathrm{e}}
\newcommand{\ii}{\mathrm{i}}
\newcommand{\dd}{\mathrm{d}}

\newcommand{\beq}[1]{\begin{equation}\label{#1}}
\newcommand{\eeq}{\end{equation}}
\newcommand{\refeq}[1]{Eq.~(\ref{#1})}

\newcommand{\refeqand}[2]{Eqs.~(\ref{#1}) and (\ref{#2})}
\newcommand{\beqm}[1]{\begin{multline}\label{#1}}

\newcommand{\refsec}[1]{Section~\ref{#1}}
\newcommand{\refsecand}[2]{Sections~\ref{#1} and \ref{#2}}
\newcommand{\refapp}[1]{Appendix~\ref{#1}}
\newcommand{\reftab}[1]{Table~\ref{#1}}
\newcommand{\reffig}[1]{Figure~\ref{#1}}
\newcommand{\reffigand}[2]{Figures~\ref{#1} and \ref{#2}}
\newcommand{\refcite}[1]{Ref.~\onlinecite{#1}}
\newcommand{\refciteand}[2]{Refs.~\onlinecite{#1} and \onlinecite{#2}}

\newcommand{\punc}[1]{\,{\text{#1}}}
\newcommand{\sub}[1]{_{\text{#1}}}
\newcommand{\zero}{^{(0)}}

\newcommand{\Ham}{H}
\newcommand{\Action}{S}
\newcommand{\Proj}{\mathcal{P}}

\newcommand{\parv}{\boldsymbol{\partial}}
\newcommand{\Dv}{\boldsymbol{D}}

\DeclareMathOperator{\Tr}{Tr}

\newcommand{\sg}[1]{\mathbb{#1}}
\newcommand{\sgQ}{\sg{Q}}
\newcommand{\sgE}{\sg{E}}
\newcommand{\sgT}{\sg{T}}
\newcommand{\sgR}{\sg{R}}
\newcommand{\sgK}{\sg{K}}

\newcommand{\msgQ}{\mathcal{Q}}
\newcommand{\msgE}{\mathcal{E}}
\newcommand{\msgT}{\mathcal{T}}
\newcommand{\msgR}{\mathcal{R}}
\newcommand{\msgK}{\mathcal{K}}

\newcommand{\spacegroup}{\mathfrak{G}}
\newcommand{\Tgroup}{\mathfrak{T}}
\newcommand{\Qgroup}{\mathfrak{Q}}

\DeclareMathOperator{\Div}{div}
\DeclareMathOperator{\Curl}{curl}
\DeclareMathOperator{\Grad}{grad}

\newcommand{\ket}[1]{\lvert#1\rangle}

\newcommand{\rv}{\boldsymbol{r}}
\newcommand{\Rv}{\boldsymbol{R}}
\newcommand{\Sv}{\boldsymbol{S}}
\newcommand{\Nv}{\boldsymbol{N}}

\newcommand{\Mv}{\boldsymbol{M}}

\newcommand{\ev}{\boldsymbol{e}}
\newcommand{\uv}{\boldsymbol{u}}
\newcommand{\deltav}{\boldsymbol{\delta}}

\newcommand{\kappav}{\boldsymbol{\kappa}}
\newcommand{\kv}{\boldsymbol{k}}

\newcommand{\sigmam}{\boldsymbol{\sigma}}
\newcommand{\thetam}{\boldsymbol{\theta}}
\newcommand{\Gammam}{\boldsymbol{\Gamma}}
\newcommand{\Deltam}{\boldsymbol{\Delta}}
\newcommand{\Dm}{\mathbf{D}}
\newcommand{\Lambdam}{\boldsymbol{\Lambda}}
\newcommand{\phiv}{\boldsymbol{\varphi}}
\newcommand{\Phiv}{\boldsymbol{\Phi}}

\newcommand{\Fv}{\boldsymbol{F}}
\newcommand{\vv}{\boldsymbol{v}}

\newcommand{\helix}{\varepsilon}

\newcommand{\alphav}{\boldsymbol{\alpha}}

\newcommand{\ProjDelta}{\hat{\Deltam}}
\newcommand{\ProjD}{\hat{\Dm}}

\newcommand{\Mm}{\mathbf{M}}
\newcommand{\MQ}{\Mm_{\sgQ}}
\newcommand{\Vv}{\boldsymbol{V}}
\newcommand{\VQ}{\Vv_{\!\!\sgQ}}

\newcommand{\sigmaQ}{\sigma_{\sgQ}}

\newcommand{\spX}{\mathrm{X}}
\newcommand{\spG}{\mathrm{\Gamma}}

\newcommand{\linkarrow}{\rightarrow}
\newcommand{\link}[2]{#1\linkarrow#2}

\newcommand{\ns}{^{\phantom{*}}}

\newcommand{\zerov}{\boldsymbol{0}}

\newcommand{\BZR}{\mathfrak{B}\sub{R}}
\newcommand{\BZL}{\mathfrak{B}\sub{L}}
\newcommand{\reducedR}[1]{[#1]}

\ifpdf

\newcommand{\putinscaledfigure}[1]{\begin{center}\includegraphics[width=\columnwidth]{#1}\end{center}}
\else

\newcommand{\putinscaledfigure}[1]{\begin{center}\includegraphics[width=\columnwidth]{#1.eps}\end{center}}
\fi

\begin{document}

\title{Higgs transitions of spin ice}

\author{Stephen Powell}
\affiliation{Joint Quantum Institute and Condensed Matter Theory Center, Department of Physics, University of Maryland, College Park, MD 20742, USA}

\begin{abstract}
Frustrated magnets such as spin ice exhibit Coulomb phases, where correlations have power-law forms at long distances. Applied perturbations can cause ordering transitions which cannot be described by the usual Landau paradigm, and are instead naturally viewed as Higgs transitions of an emergent gauge theory. Starting from a classical statistical model of spin ice, it is shown that a variety of possible phases and transitions can be described by this approach. Certain cases are identified where continuous transitions are argued to be likely; the predicted critical behavior may be tested in experiments or numerical simulations.
\end{abstract}


\maketitle

\section{Introduction}
\label{SecIntroduction}

The frustrated magnetic compounds known as spin ice,\cite{Harris,Bramwell} including $\mathrm{Ho}_2\mathrm{Ti}_2\mathrm{O}_7$ and $\mathrm{Dy}_2\mathrm{Ti}_2\mathrm{O}_7$, are well described by a classical Ising model on the pyrochlore lattice. As originally noted by Anderson,\cite{Anderson2} this model has an extensively degenerate low-energy manifold closely related to that of water ice;\cite{Pauling} experiments\cite{Ramirez} indeed observe the characteristic low-temperature residual entropy. Neutron scattering reveals long-ranged dipolar correlations in these materials, which are in close agreement with numerical simulations.\cite{Fennell2,MelkoReview,Fennell}

Such power-law correlations are a signature of the ``Coulomb phase''\cite{HenleyReview} common in frustrated systems, where the low-energy manifold is highly degenerate but strongly constrained. Models such as spin ice, where degrees of freedom on the links of a lattice are constrained to have zero divergence (see \refcite{HenleyReview} and \refsec{SecGaugeTheory}), can be described by a lattice gauge theory with magnetic monopoles\cite{Castelnovo} forbidden. The dipolar correlations can be understood by proceeding to a long-wavelength description with the link variables replaced by a coarse-grained vector field.\cite{Youngblood,Huse,Henley,Isakov1}

Starting from a nearest-neighbor model of spin ice,\cite{Isakov2} applying perturbations (for example, magnetic field,\cite{Moessner111,Yoshida,Jaubert1,SpinIceCQ} pressure,\cite{Jaubert2} impurities,\cite{Andreanov} or simply the residual further-neighbor dipole interaction\cite{Bramwell,Melko}) can lead to transitions at sufficiently low temperature. The resulting ordered phases, which have a finite correlation length and structure depending on microscopic details, cannot be described by the coarse-grained theory of the Coulomb phase. (The effectively discrete degrees of freedom in spin ice should be contrasted with $\mathrm{O}(N)$ spin systems, where more conventional critical theories have been derived from a coarse-grained theory.\cite{Pickles,Xu})

In the present work, these ordering transitions are analysed starting from a noncompact $\mathrm{U}(1)$ lattice gauge theory, in terms of which they are (Anderson\cite{Anderson1}--)Higgs transitions involving condensation of an emergent matter field (dual to the magnetic monopoles). This perspective leads to a long-wavelength description capturing both the Coulomb and ordered phases, which plays the role of the Ginzburg-Landau theory for conventional ordering transitions.\cite{Landau} This description is strongly constrained by the symmetries, physical and emergent, of the gauge theory, leading to nontrivial predictions for the allowed ordered states and transitions.

Of particular theoretical interest are potential continuous Higgs transitions, which belong to the class of transitions beyond the Landau-Ginzburg-Wilson paradigm that has recently been the subject of considerable attention.\cite{Motrunich,Senthil,Balents} Numerical simulations are better suited to determining the order of a given transition, but the focus here will be on cases where continuous transitions appear most plausible. These are argued to include transitions into the spin spiral illustrated in \reffig{FigSpinSpiral} and the bond-ordered paramagnet of \reffig{FigBondOrdered}; see \refsec{SecConclusions} for a summary of the transitions. The resulting critical behavior may be observable in experiment, but the particular forms of the applied perturbations and the simultaneous requirements of low monopole density and ergodicity mean that testing these predictions is likely more feasible using numerical simulations.

Previous studies of phase transitions in spin ice have considered the low-temperature ordered states favored by the dipolar interactions,\cite{Melko}  magnetic fields,\cite{Moessner111,Yoshida,Jaubert1,SpinIceCQ} and pressure.\cite{Jaubert2} While such transitions can be included in the framework described here, the present work excludes states with nonzero net (spin) polarization (see \refsec{SecGaugeTheory}); these include those favored by application of a uniform magnetic field.\cite{Moessner111,Yoshida,Jaubert1,SpinIceCQ} Such states require a generalization of the method, which will be presented elsewhere.

The present analysis has connections with studies of the cubic dimer model\cite{Alet,Charrier,CubicDimers,Chen,Papanikolaou,Charrier2} and quantum models on the pyrochlore lattice,\cite{Bergman} both of which exhibit Higgs transitions. An interesting feature of spin ice, distinguishing it from these precedents, is symmetry under spin inversion, to be referred to as ``time reversal''. As shown in \refsecand{SecTimeReversal1}{SecTimeReversal2}, this causes extra ``Kramers'' degeneracies, manifested in the hopping model studied in \refapp{AppHoppingModel} as Dirac cones in the spectrum (see \reffig{FigHoppingDispersion}). These imply that the matter modes of relevance to the Higgs transitions are more numerous than in cases studied previously, leading to a greater diversity of ordered phases and transitions.

This work will treat a classical statistical model, appropriate for spin ice over a range of temperatures, but is also of relevance to quantum systems. The long-wavelength description of these three-dimensional (3D) classical transitions is identical to certain quantum phase transitions in 2D systems.\cite{Sachdev} (It is in some cases possible to make the relationship explicit by performing the mapping in terms of microscopic models.\cite{Jaubert1,SpinIceCQ,CubicDimers}) A second connection is that 3D quantum models with phase transitions at nonzero temperature can be described by classical actions;\cite{Bergman} these include transitions in spin ice where quantum effects lead to ordered phases.\cite{Shannon} The classical physics of spin ice is also of relevance to the quantum Heisenberg model on the pyrochlore lattice, where arrangements of singlet dimers can be mapped to spin-ice configurations.\cite{Nussinov}

\subsection{Outline}
\label{SecOutline}

The route from the microscopic model, introduced in \refsec{SecSpinIce}, to a field-theoretical description is chosen to emphasize the nature of the Higgs transitions. The mapping of \refsec{SecGaugeTheory} leads to a lattice gauge theory containing a matter field defined on the sites of a diamond lattice, and the transitions of interest occur when certain modes of this field condense. Because this is a strongly interacting model, these cannot be determined exactly but can be labeled by the irreducible representations of the symmetry group. The symmetries, discussed in \refsec{SecSymmetries}, are modified in an essential way by the presence of the gauge field; the representations themselves are treated in \refsec{SecRepresentations}.

In \refsec{SecOrderParameters}, the physical observables of the spin model are related to the fields appearing in the gauge theory, using the standard procedure of identifying quantities with identical behavior under the symmetries. The Higgs transition is naturally described in terms of the matter fields, and it will be shown that order parameters for certain phases can expressed as functions of these. Once a physical quantity is associated with each bilinear, one can construct an expansion, similar to a Ginzburg-Landau theory, in powers of the fields and derivatives, involving all terms allowed by symmetry. (In contrast to the standard Landau paradigm, the primitive fields that appear are not the order parameters for the transition, but instead emergent matter and gauge fields.) This expansion is found in \refsec{SecPhaseTransitions}, allowing the ordered states and transitions to be determined. A summary of the important results and conclusions is given in \refsec{SecConclusions}.

\subsection{Spin ice}
\label{SecSpinIce}

\begin{figure}
\putinscaledfigure{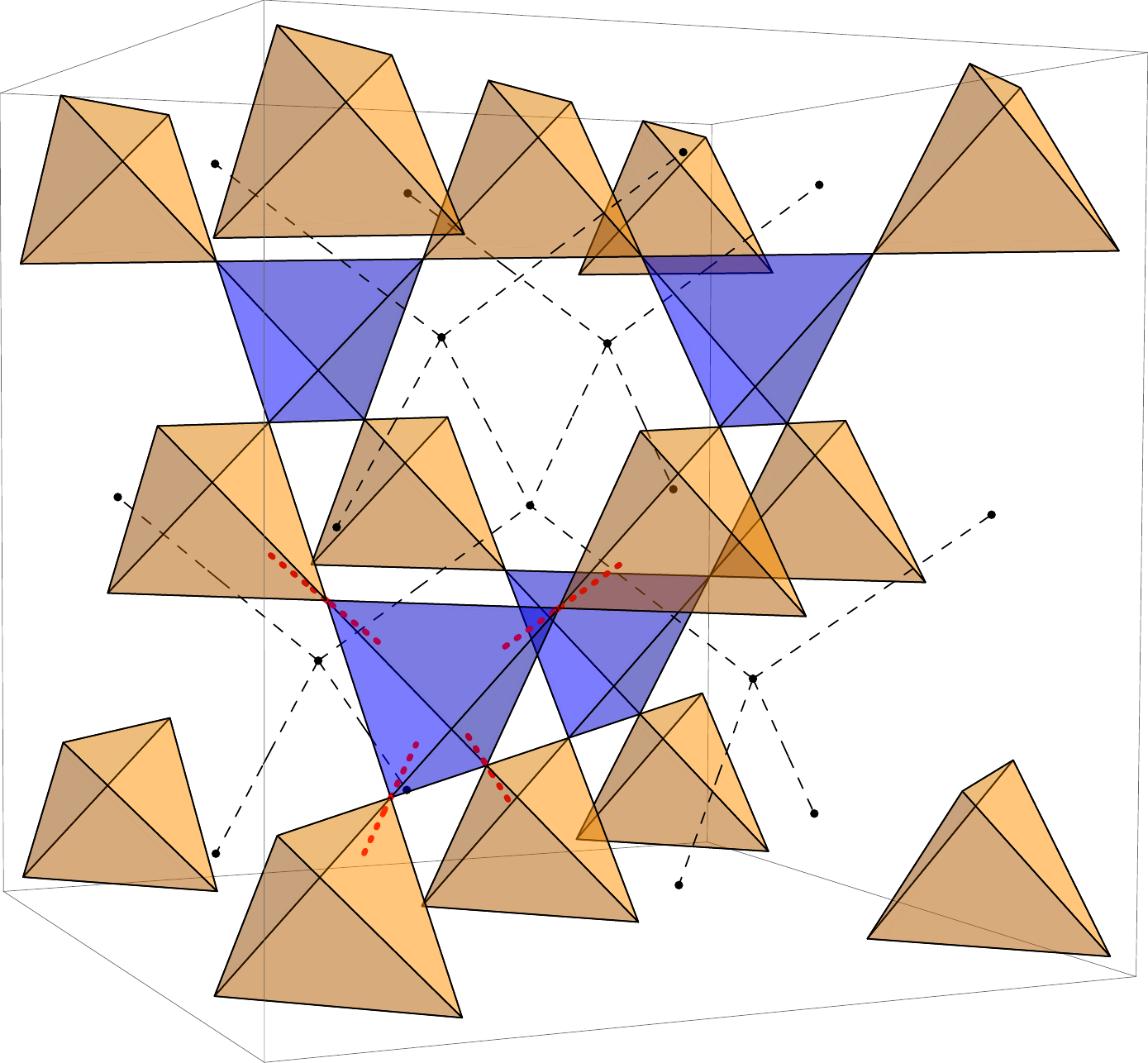}
\caption{The pyrochlore lattice shown as a network of corner-sharing tetrahedra (orange and blue), along with the dual diamond lattice (dashed lines). In spin ice, magnetic moments are located at the sites of pyrochlore, which are at the vertices of the tetrahedra. A strong easy-axis anisotropy requires the spins to align with the axes connecting the centers of neighboring tetrahedra; dotted (red) lines illustrate these for the sites of a single tetrahedron.
\label{FigPyrochloreDiamond}}
\end{figure}

``Spin ice''\cite{Harris,Bramwell} refers to a class of frustrated magnetic materials, such as $\mathrm{Ho}_2\mathrm{Ti}_2\mathrm{O}_7$ and $\mathrm{Dy}_2\mathrm{Ti}_2\mathrm{O}_7$, with local moments $\Sv$ on the sites $\rv$ of a pyrochlore lattice, shown in \reffig{FigPyrochloreDiamond}. In these materials, the magnetic moments are large ($\sim 10$ Bohr magnetons) and experience strong local easy-axis crystal fields, so an accurate microscopic description is provided by a classical model of Ising spins with a combination of exchange and long-range dipolar interactions.\cite{Bramwell,MelkoReview} The behavior on length scales larger than a few interatomic spacings is nonetheless accurately described by a model with purely nearest-neighbor ferromagnetic interactions, as a result of a remarkable ``projective equivalence''.\cite{Isakov2,Castelnovo} Specifically, a nearest-neighbor model has an extensively degenerate minimal-energy manifold that is identical to the set of low-energy states of the full interaction potential, known from comparison of experiment with simulations.\cite{Bramwell,Fennell2,MelkoReview,Fennell} (The small energy differences between these states in the latter case are, along with quantum effects, negligible for a range of temperatures.)

With this is mind, one can write the configuration energy as
\beq{EqSpinIceHamiltonian}
\Ham = -J\sum_{\langle\rv,\rv'\rangle} \Sv_{\rv}\cdot \Sv_{\rv'} + V(\Sv)\punc{,}
\eeq
where $J > 0$ and the local crystal fields constrain the spins to $\Sv_{\rv} = s_{\rv} \hat\deltav_{\rv}$, with $s_{\rv} = \pm 1$. The static unit vector $\hat\deltav_{\rv}$ lies along the axis joining the centers of the two tetrahedra that share the site $\rv$, as shown in \reffig{FigPyrochloreDiamond}, and can be chosen always to point outwards from tetrahedra of a given orientation. The term $V(\Sv)$ is a perturbation that may represent coupling to an external field or additional interactions between spins (such as those beyond nearest neighbors). Specific examples will be given in \refsec{SecPhaseTransitions}; these include cases where $V$ preserves the full symmetry of the pyrochlore lattice as well as the Ising symmetry under $s_{\rv} \leftarrow -s_{\rv}$, and those where it breaks certain symmetries. Most of the analysis will require only that $\lvert V\rvert \ll J$, where $\lvert V\rvert$ denotes the scale of couplings in $V$.

The first term of \refeq{EqSpinIceHamiltonian} can be rewritten, up to a constant, as
\beq{EqHamJ}
\Ham^{(J)} = -\frac{J}{2}\sum_{i} \left\lvert\sum_{\rv \in i} \Sv_{\rv}\right\rvert^2\punc{,}
\eeq
where the outer sum is over all tetrahedra and $\rv \in i$ denotes all sites in tetrahedron $i$. Inspection of a single tetrahedron shows that there are $6$ configurations for its spins that maximize $\lvert\sum \Sv_{\rv}\rvert$, with the vector sum aligned with one of the $6$ cubic directions. In the pyrochlore lattice, neighboring tetrahedra are constrained by their shared spin, so the total number of allowed configurations is less than $6^{N/2}$ (for a lattice with $N$ spins), but still grows exponentially with $N$ and is in fact closely approximated by Pauling's estimate\cite{Pauling} of $(3/2)^{N/2}$.

In the present work, it will be assumed throughout that the temperature $T \sim \lvert V\rvert \ll J$, so the system is constrained to states within this minimal-energy manifold. These are said to obey the ``ice rule'': a tetrahedron with total spin maximal must have two spins pointing inward and two pointing outward. In the absence of the perturbation $V$, the partition function is given in this limit by a sum with equal weight over all ice-rule configurations. It is well established, based on theoretical arguments and numerics,\cite{HenleyReview,Youngblood,Huse,Henley,Isakov1} that such a sum implies long-range correlations between the spins of the form
\beq{EqCoulombCorrelations}
\langle S_{\rv}^\mu S_{\zerov}^\nu \rangle \propto \frac{3r_\mu r_\nu - \lvert\rv\rvert^2\delta_{\mu\nu}}{\lvert\rv\rvert^5}
\punc{.}
\eeq
These are the signature of a ``Coulomb phase'', and can be found using an effective coarse-grained description in terms of a gauge theory.\cite{HenleyReview} Such correlations have indeed been observed in neutron-scattering experiments.\cite{Fennell2,MelkoReview,Fennell}

The perturbation $V$ acts to distinguish the ice-rule states (i.e., give them different Boltzmann weights), and so is effectively projected to this manifold. For $\lvert V\rvert \gtrsim T$ it can lead to an ordered phase with a finite correlation length, and it is the transitions into these phases that will be the subject of this work. Although the transition may involve the formation of symmetry-breaking order, a standard Landau theory written in terms of an order parameter is not applicable, since it fails to account for the Coulomb correlations on the high-temperature side.\cite{Fisher} (Here, ``high-temperature'' means $\lvert V\rvert \lesssim T \ll J$; for $T \sim J$ the system is a conventional paramagnet.) Such transitions can instead be viewed as Higgs transitions in the effective gauge theory, and this perspective allows a description that captures both the Coulomb-phase correlations and the microscopic details that are crucial to an ordered state.

\section{Gauge theory}
\label{SecGaugeTheory}

\begin{figure}
\putinscaledfigure{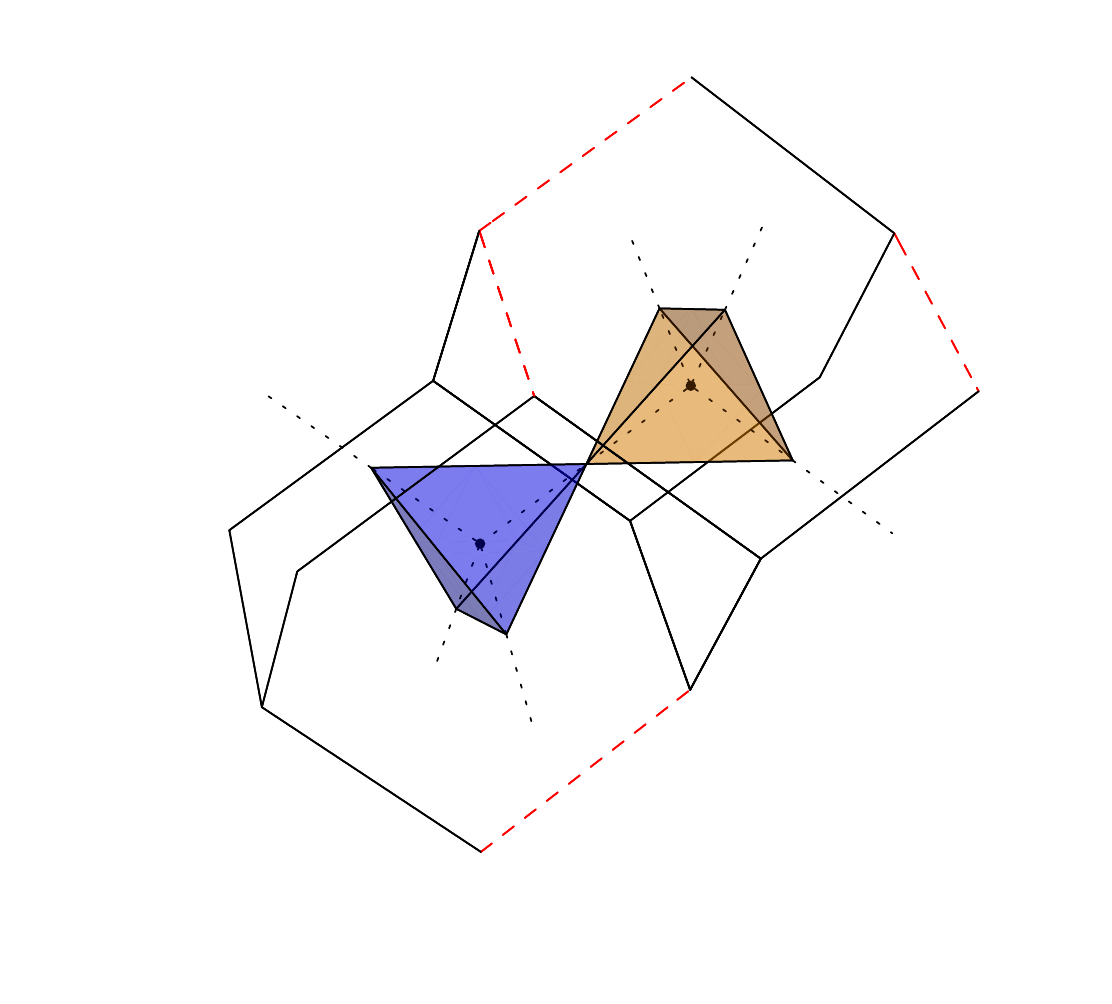}
\caption{Two tetrahedra of the pyrochlore lattice (orange and blue), along with the diamond sites (points) situated at the centers of each tetrahedron. Each site of pyrochlore, equivalent to a diamond link, is surrounded by a (nonplanar) hexagonal plaquette of the dual diamond lattice (solid black and dashed red lines), so a tetrahedron is enclosed in a ``cage'' consisting of $4$ hexagons and $10$ diamond sites. The dual-diamond links $\lambda$ on which the static gauge field $a_\lambda$ takes the value $\frac{1}{2}$ are shown as dashed (red) lines; the remainder have $a_\lambda = 0$. Every plaquette has an odd number of links with $a_\lambda \neq 0$ and so net flux $\frac{1}{2}$ (up to an integer), as required by \refeq{EqDefinea}. (This choice of $a_\lambda$ is shown for a larger region of the dual diamond lattice in \reffig{FigGaugeFieldDiagram}.)
\label{FigPyrochloreDiamond2}}
\end{figure}

The Coulomb correlation functions of \refeq{EqCoulombCorrelations} can be derived by describing the spin model in terms of a coarse-grained gauge theory.\cite{HenleyReview} In order to preserve those details of the microscopics that are crucial to the ordering transitions, this section will instead present a mapping to a gauge theory on the lattice. As illustrated in \reffig{FigPyrochloreDiamond2}, the sites $\rv$ of pyrochlore are equivalent to the links of a diamond lattice (pyrochlore is the ``medial lattice'' of diamond\cite{HenleyReview}). For each directed link $\ell$, define $B_{\ell}$ taking the value $\pm \frac{1}{2}$ if the corresponding spin $\Sv_{\rv}$ is aligned parallel or antiparallel to the link.

When $T \ll J$, only states within the ice-rule manifold are accessible: the spins are arranged so that two spins point into and two out of every tetrahedron of the pyrochlore lattice. Such an arrangement obeys the simple condition
\beq{EqDivB}
\Div_i B = 0\punc{,}
\eeq
where $\Div_i$ denotes the lattice divergence at the diamond site $i$, given by the sum of $B_\ell$ on links $\ell$ pointing outwards from site $i$. This description in terms of an effective magnetic field provides an immediate motivation for the gauge-theoretical description, and upon coarse-graining\cite{HenleyReview} implies the correlation function of \refeq{EqCoulombCorrelations}.

Since this work will address the case when the ice rule is strictly enforced, \refeq{EqDivB} will be treated as a constraint, and the configuration energy is given simply by
\beq{EqEB}
\Ham_B = V(B)\punc{,}
\eeq
where the function $V$ appearing in \refeq{EqSpinIceHamiltonian} has been rewritten in terms of $B$. The partition function is given by summing over all configurations of $B$ consistent with \refeq{EqDivB} with Boltzmann weight $\ee^{-\Ham_B/T}$.

To resolve the constraint of \refeq{EqDivB}, one can express $B$ in terms of a vector potential. The diamond lattice is self-dual, so define $A$ on the (directed) links $\lambda$ of a dual diamond lattice by
\beq{EqBfromA}
B_\ell = \Curl_{\pi(\ell)} A\punc{,}
\eeq
where $\Curl_{\pi(\ell)}$ is the lattice curl around the plaquette (nonplanar hexagon) $\pi(\ell)$ dual to the link $\ell$ (see \reffig{FigPyrochloreDiamond2}).

Expressing $B$ as in \refeq{EqBfromA} automatically incorporates the constraint on its divergence, but to rewrite the problem in terms of $A$, one must also allow for the restriction of $B_\ell$ to the values $\pm \frac{1}{2}$ on each link $\ell$. This will be done in a way that may appear unnecessarily complicated, but that makes the nature of the Higgs transition directly apparent. One can allow $A_\lambda$ to take all real values by adding constraining terms to the energy,
\begin{multline}
\label{EqEA1}
\Ham_A = -\Lambda \sum_\lambda \cos 2\pi(A_\lambda - a_\lambda) + U\sum_\pi (\Curl_\pi A)^2 \\{}+ V(\Curl A)\punc{.}
\end{multline}
The first term involves the static background $a$, which is defined so that
\beq{EqDefinea}
\Curl_\pi a = \frac{1}{2}
\eeq
up to an integer, for all dual plaquettes $\pi$. For large $\Lambda$, the first term in \refeq{EqEA1} ensures that $A_\lambda$ differs from $a_\lambda$ by an integer on every link $\lambda$, and hence that $\Curl_\pi A$ is a half-integer for every plaquette $\pi$. The second term then selects the values $\pm \frac{1}{2}$, so that $\Ham_A$ reproduces the original model when the parameters obey $\Lambda \gg U \gg \lvert V\rvert, T$. The partition function is given by integrating over $A_\lambda$ on every link $\lambda$ and is identical (in the appropriate limit) to that given by a sum over configurations of $B$.

Any choice may be made for the background field $a$, provided that \refeq{EqDefinea} is satisfied. This expression clearly requires that $a$ be nonuniform, and it in fact has important implications for symmetry, as discussed in \refsec{SecSymmetries}. (There is a close relation with the Hofstadter problem,\cite{Hofstadter} which is explored in more detail in \refapp{AppHoppingModel}.) \reffigand{FigPyrochloreDiamond2}{FigGaugeFieldDiagram} show a particular choice that obeys \refeq{EqDefinea} and has the smallest possible unit cell, containing two unit cells of the diamond lattice.
\begin{figure}
\putinscaledfigure{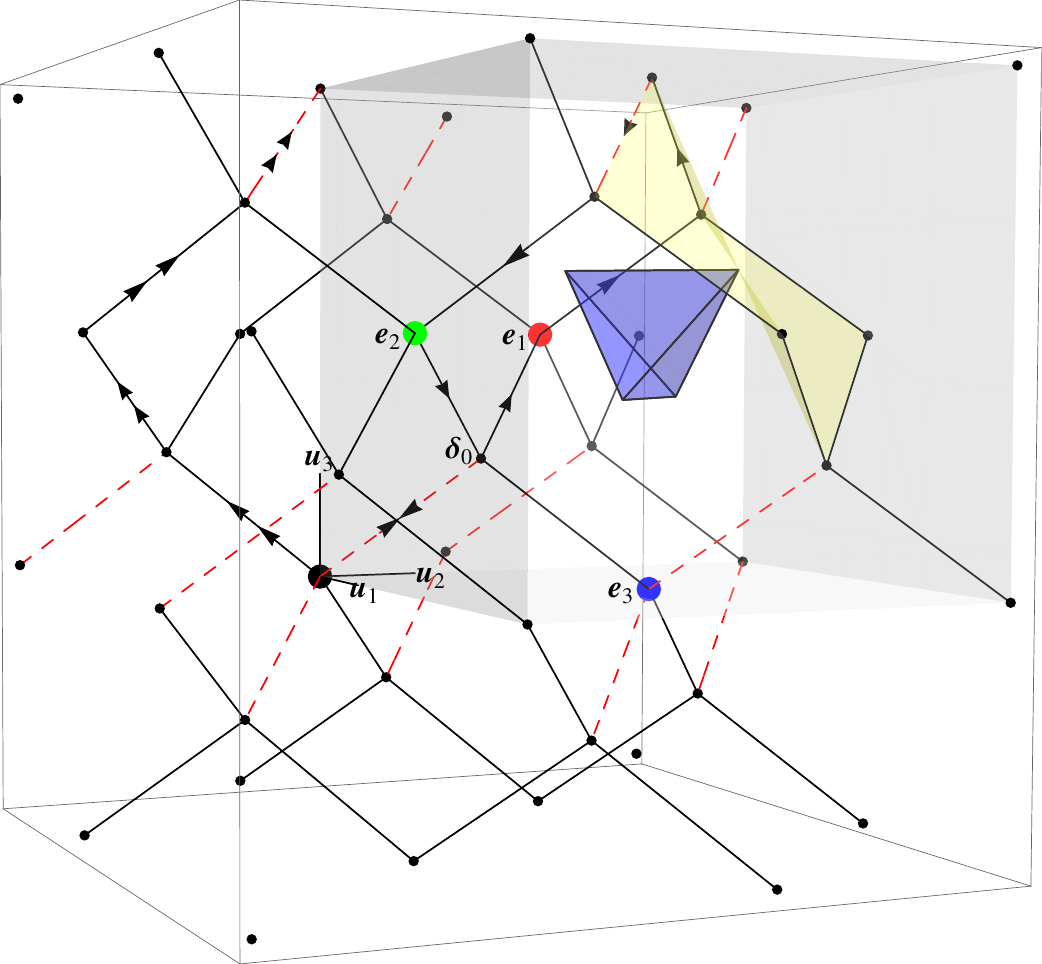}
\caption{The dual diamond lattice, showing a configuration of the static gauge field $a_\lambda$ (solid black and dashed red lines, as in \reffig{FigPyrochloreDiamond2}) that obeys \refeq{EqDefinea}. A hexagonal plaquette and one tetrahedron of the pyrochlore lattice are also shown (in yellow and blue respectively); see \reffig{FigPyrochloreDiamond}. The Cartesian coordinate axes are labeled $\uv_i$, and the large points show the origin of the coordinate system (black) and the three unit vectors $\ev_j$ (red, green, blue) of face-centered cubic (FCC). Diamond consists of this Bravais lattice decorated with a two-site unit cell, with sites at $\zerov$ and $\deltav_0$. The large shaded cube is the unit cell of the simple cubic lattice corresponding to the FCC lattice. Arrows on certain diamond links indicate the chains appearing in \refeqand{EqscTexpand}{EqphiK} (single and double arrowheads respectively).
\label{FigGaugeFieldDiagram}}
\end{figure}

The definition of $A$ in \refeq{EqBfromA} implies that there is a redundancy under
\beq{EqGaugeSymmetry1}
A_\lambda \leftarrow A_\lambda + \Grad_\lambda X\punc{,}
\eeq
where $X$ is defined on the (dual) sites $j$, and $\Grad$ is the lattice gradient. The constraints in \refeq{EqEA1} imply that this gauge symmetry is discrete, given fixed background $a$, despite the continuous nature of $A$. (The redundancy also means that any choice of background obeying \refeq{EqDefinea} is equivalent.)

This gauge symmetry can be made continuous by introducing new degrees of freedom $\theta_j$, defined on the sites $j$ of the dual diamond lattice, that are ``minimally coupled'' to the gauge field $A$. The new energy function is
\begin{multline}
\label{EqEA2}
\Ham_{A,\theta} = -\Lambda \sum_\lambda \cos[2\pi(A_\lambda - a_\lambda) - \Grad_\lambda \theta] \\
{}+ U \sum_\pi (\Curl_\pi A)^2 + V(\Curl A)\punc{,}
\end{multline}
with a continuous gauge symmetry
\beq{EqGaugeSymmetry2}
\begin{aligned}
A_\lambda &\leftarrow A_\lambda + \Grad_\lambda X\\
\theta_j &\leftarrow \theta_j + 2\pi X_j\punc{.}
\end{aligned}
\eeq
The latter expression implies that $\theta$ is redundant and its introduction does not change the value of the partition function.

Despite this redundancy, it is convenient to include the ``matter field'' $\theta$, giving an XY model coupled to a $\mathrm{U}(1)$ gauge field.\cite{EndNoteCharges} A Higgs transition occurs when $T$ is decreased and the matter field $\theta$ ``condenses'', leading to a quadratic term in the effective action for the gauge field $A$ and the loss of the power-law correlations of the Coulomb phase, \refeq{EqCoulombCorrelations}. The background field $a$, which implements certain microscopic details of the original model, implies that the condensate configuration is nonuniform. The Higgs transition therefore also spontaneously breaks the spatial symmetry, and the spatial order is manifest in gauge-invariant combinations of the fields $\theta$ and $A$.

A description of the Higgs transition is given in terms of those modes of the matter field that are most important near the transition. One can eliminate (``integrate out'') others, renormalizing the parameters appearing in the action. This will have the effect of softening the constraints enforced by the parameters $\Lambda$ and $U$ in \refeq{EqEA2}, without modifying the phase structure or the universal behavior near any continuous phase transitions. Similarly, the matter field will be replaced by a complex variable $\psi_j \sim \ee^{\ii \theta_j}$, whose modulus will not be constrained. In these terms, \refeq{EqEA2} is a classical analogue of bosons hopping on a lattice; the latter model is studied in the \refapp{AppHoppingModel}. (It should be noted, however, that the lattice is three-dimensional, and so this is not the classical--quantum mapping used previously to study transitions in spin ice.\cite{Jaubert1,SpinIceCQ})\label{SecQuantumAnalogy}

The precise form of the relevant modes cannot be determined analytically in this strongly coupled system, but it is possible to categorize them in terms of representations of the symmetry group. This will involve a treatment of the modification of the spatial symmetries due to the background field $a$, with which the next section will be concerned.

While \refeq{EqBfromA} can describe any configuration obeying the constraint \refeq{EqDivB}, it is in fact inconvenient in the case of a nonzero total spin polarization. It is then preferable to replace \refeq{EqBfromA} by $B_\ell = b_\ell + \Curl_{\pi(\ell)} A$, where $b$ is a static magnetic field. The latter quantity is conserved by any local dynamics within the ice-rule manifold and acts as an additional flux on the matter fields $\psi_j$, modifying the analysis that follows. The present work will exclude states with nonzero overall polarization in any direction; perturbations that favor them, such as uniform magnetic fields, will be addressed with this approach elsewhere.

\section{Symmetries}
\label{SecSymmetries}

As for more conventional Ginzburg-Landau theories, symmetries provide an important tool in constructing the long-wavelength action. In the present case, the modifications due to the background gauge field $a$ are particularly important. These reflect details of the microscopic physics that can be neglected in the coarse-graining procedure appropriate to the Coulomb phase,\cite{Youngblood,Huse,Henley,Isakov1} but are crucial for ordering transitions.

\subsection{Lattice symmetries}
\label{SecLatticeSymmetries}

The effective model that will be treated in the following sections is $\Ham_{A,\theta}$, defined in \refeq{EqEA2}, whose degrees of freedom occupy the sites and links of a diamond lattice. The discussion of the spatial symmetries will therefore be couched in terms of diamond, but the same symmetries apply to pyrochlore, its medial lattice. (It should be recalled from \refsec{SecGaugeTheory} that the diamond lattice on which $\theta$ is defined is dual to the original diamond lattice, whose links correspond to pyrochlore sites.)

The Bravais lattice of diamond (see \reffig{FigGaugeFieldDiagram}) is face-centered cubic (FCC), with primitive vectors $\ev_i$ for $i \in \{1,2,3\}$, related to the Cartesian unit vectors $\uv_i$ by $\ev_1 = 2(\uv_2 + \uv_3)$ and cyclic permutations. (In these units, the cubic unit cell of FCC has side $4$ and the nearest-neighbor distance in the pyrochlore lattice is $\sqrt{2}$.) Translations by the lattice vectors $\ev_i$ are effected by the operators $\sgT_i$. In the pyrochlore lattice, these translations map a tetrahedron onto another of the same orientation.

The diamond lattice has two sites within the unit cell, labeled by sublattice $\sigma \in \{0,1\}$ and separated by displacement $\deltav_0 = \uv_1 + \uv_2 + \uv_3 = \frac{1}{4}(\ev_1 + \ev_2 + \ev_3)$. (The two sites correspond to tetrahedra of opposite orientation in the pyrochlore lattice.) A site on sublattice $\sigma$ has $4$ nearest neighbors, all on the opposite sublattice, at displacements $(-1)^{\sigma}\deltav_\mu$, for $\mu \in \{0,\ldots,3\}$, where $\deltav_i = \deltav_0 - \ev_i$.

The crystallographic space group $\spacegroup$ of the diamond lattice\cite{Cornwell} is denoted $O_h^7$, while the corresponding point group is $O_h$. In other words, if the action of each lattice symmetry $\sgQ$ on a position vector $\rv$ is given by $\rv \rightarrow \sgQ\rv = \MQ \rv + \VQ$, then the rotational parts $\MQ$ form the group $O_h$. As for any space group, the pure translations $\sgT$ form an invariant subgroup $\Tgroup$, and any element of $\spacegroup$ can be constructed as the product $\sgT\sgQ$ of some $\sgT \in \Tgroup$ and one of the ``coset representatives'' $\sgQ \in \Qgroup$. The elements of $\Qgroup$ can be found by associating with each element $\MQ$ of $O_h$ the appropriate translation $\VQ$, unique up to a lattice vector, that maps the lattice onto itself.\cite{Cornwell} Because of the two-site unit cell, the diamond lattice is nonsymmorphic (i.e., $O_h \not\subset O_h^7$), having $\VQ = \deltav_0$ for certain elements of $O_h$. For such transformations, which exchange the two diamond sublattices (or two orientations of tetrahedra in the pyrochlore lattice), let $\sigmaQ = 1$; otherwise $\sigmaQ = 0$. The $10$ classes of $O_h$ are listed in \reftab{TabSymmetries}, along with a description of the corresponding transformations of the diamond lattice.
\begin{table*}
\caption{\label{TabSymmetries}Classes of the point group $O_h$,\cite{Cornwell} and corresponding symmetries $\sgQ$ of the diamond and pyrochlore lattices. For each class, $\det \sgQ = \pm 1$ for proper and improper transformations respectively, and $\sigmaQ = 1$ for those that exchange diamond sublattices (or pyrochlore tetrahedra of opposite orientations). In each case an example element is listed using ``XYZ'' notation:\cite{Lax} the order of the letters indicates the permutation given by the matrix $\MQ$ and bars signify minus signs.}
\begin{ruledtabular}
\begin{tabular}{lccccl}
Class & Order & $\det \sgQ$ & $\sigmaQ$ & Example & Description \\\hline
$E$ & $1$ & $1$ & $0$ & $XYZ$ & Identity ($\sgE$)\\
$8C_3$ & $8$ & $1$ & $0$ & $YZX$ & Proper rotation ($\sgR$) by $\pm\frac{2\pi}{3}$ about diamond link $\deltav_i$\\
$3C_2$ & $3$ & $1$ & $0$ & $X\bar{Y}\bar{Z}$ & Proper rotation by $\pi$ about cubic axis $\uv_i$\\
$6C_4$ & $6$ & $1$ & $1$ & $XZ\bar{Y}$ & Screw rotation by $\pm \frac{\pi}{2}$ about cubic axis $\uv_i$\\
$6C_2'$ & $6$ & $1$ & $1$ & $\bar{X}ZY$ & Screw rotation by $\pi$ about FCC vector $\ev_i$\\
$I$ & $1$ & $-1$ & $1$ & $\bar{X}\bar{Y}\bar{Z}$ & Inversion\\
$8S_6$ & $8$ & $-1$ & $1$ & $\bar{Y}\bar{Z}\bar{X}$ & Improper rotation by $\pm\frac{2\pi}{3}$ about diamond link $\deltav_i$\\
$3\sigma_h$ & $3$ & $-1$ & $1$ & $\bar{X} Y Z$ & Glide reflection in plane perpendicular to $\uv_i$\\
$6S_4$ & $6$ & $-1$ & $0$ & $\bar{X} \bar{Z} Y$ & Improper rotation by $\pm \frac{\pi}{2}$ about cubic axis $\uv_i$\\
$6\sigma_d$ & $6$ & $-1$ & $0$ & $X\bar{Z}\bar{Y}$ & Reflection in plane normal to FCC vector $\ev_i$
\end{tabular}
\end{ruledtabular}
\end{table*}

Perturbations such as an applied magnetic field or uniaxial pressure can reduce the symmetry to a subgroup of $\spacegroup$. This will modify the couplings allowed to appear in the long-wavelength description and lead to different classes of transitions and critical behavior, as discussed in \refsec{SecPhaseTransitions}.

\subsection{Magnetic symmetries}
\label{SecMSG}

With any choice of the background gauge potential $a$, the fields $A$ and $\theta$ (or $\psi$) obey a modified set of symmetries, which will be referred to as the ``magnetic symmetry group'' (MSG).\cite{Brown,Zak,Overhof,EndNotePSG} For each element $\sgQ \in \spacegroup$, there is a corresponding gauge transformation that must accompany $\sgQ$ to give a symmetry of \refeq{EqEA2}.

The first term of $\Ham_{A,\theta}$ (other terms are trivially symmetric) can be rewritten in terms of $\psi_j = \ee^{\ii \theta_j}$ as
\beq{EqEA3}
\Ham^{(\Lambda)} = -\frac{\Lambda}{2} \sum_{\link i j} \ee^{2\pi\ii(A_{\link i j} - a_{\link i j})}\psi^*_j\psi\ns_i\punc{,}
\eeq
where the sum runs over directed links $\link i j$ of the dual diamond lattice, with each link counted twice. Under a symmetry $\sgQ$ that maps the site $j$ to $\sgQ j$, the degrees of freedom transform according to
\beq{EqApsisgQ}
\begin{aligned}
\psi_j &\xleftarrow{\sgQ} \psi_{\sgQ^{-1}j}\\
A_{\link i j} &\xleftarrow{\sgQ} A_{\link{\sgQ^{-1}i}{\sgQ^{-1}j}}\punc{,}
\end{aligned}
\eeq
so \refeq{EqEA3} becomes
\beq{EqEA4}
\Ham^{(\Lambda)} = -\frac{\Lambda}{2} \sum_{\link i j} \ee^{2\pi\ii(A_{\link i j} - a_{\link {\sgQ i} {\sgQ j}})} \psi_j^*\psi_i\ns\punc{,}
\eeq
after a change of the summation variables.

The transformation $\sgQ$ is therefore not by itself a symmetry of $\Ham^{(\Lambda)}$, but must be accompanied by a gauge transformation. For every $\sgQ\in\spacegroup$, define a corresponding operator $\msgQ$ under which
\beq{EqDefinemsgQ}
\begin{aligned}
\psi_j &\xleftarrow{\msgQ} \pi_{\msgQ}(\sgQ^{-1}j)\psi_{\sgQ^{-1}j}\\
A_{\link i j} &\xleftarrow{\msgQ} A_{\link{\sgQ^{-1}i}{\sgQ^{-1}j}}\punc{,}
\end{aligned}
\eeq
where the phase factor $\pi_{\msgQ}$ compensates the change in background field induced by $\sgQ$. Comparison of \refeqand{EqEA3}{EqEA4} shows that one must choose $\pi_{\msgQ}$ so that
\beq{EqDefinepi}
\pi_{\msgQ}^*(j)\pi_{\msgQ}\ns(i) = \omega^*(\link ij) \omega(\link{\sgQ i}{\sgQ j})\punc{,}
\eeq
where $i$ and $j$ are neighboring sites and $\omega(\lambda) = \ee^{2\pi\ii a_{\lambda}}$. Given a fixed choice of background field, the defining equation \refeq{EqDefinepi} specifies $\pi_{\msgQ}(j)$ for all sites $j$, up to an arbitrary uniform phase factor. Note that both $\pi_{\msgQ}$ and $\omega$ are phase factors, so $\pi_{\msgQ}^*(j) = \pi_{\msgQ}^{-1}(j)$ and $\omega^*(\link ij) = \omega^{-1}(\link ij) = \omega(\link ji)$. 

For the particular case of \refeq{EqDefinea}, it is in fact possible to choose the background field so that $a_\lambda$ takes the values $0$, $+\frac{1}{2}$ and $-\frac{1}{2}$ on every link $\lambda$, as illustrated in \reffig{FigGaugeFieldDiagram}. In this case $\omega$ and $\pi_{\msgQ}$ are both real, but this assumption does not lead to any significant simplification and will not be made in the following. (Note that the fact that a real gauge exists implies that all gauge-invariant quantities are real.)

One can extend the definition of $\omega$ to chains of links,
\beq{EqomegaChain}
\omega(i\linkarrow j \linkarrow \cdots \linkarrow \ell) = \omega(i \linkarrow j)\omega(j \linkarrow \cdots \linkarrow \ell)\punc{,}
\eeq
allowing \refeq{EqDefinepi} to be generalized to
\beq{EqDefinepi2}
\pi_{\msgQ}^*(j)\pi_{\msgQ}\ns(i) = \omega^*(i \linkarrow \cdots \linkarrow j) \omega({\sgQ i} \linkarrow \cdots \linkarrow {\sgQ j})\punc{,}
\eeq
where the sites $i$ and $j$ need not be neighbors. The sites visited by the two chains in \refeq{EqDefinepi2} must be in correspondence, but the path is otherwise arbitrary.

It should be noted that $\omega(i \linkarrow \cdots \linkarrow j)$ depends on the route taken by the chain, because the background field $a$ has nonzero lattice curl. In fact, because $\Curl a = \pm \frac{1}{2}$ for all hexagonal plaquettes of the dual diamond lattice, any closed path has $\omega(i\linkarrow \cdots \linkarrow i) = (-1)^{N\sub{h}}$ where $N\sub{h}$ is the number of plaquettes it encloses. This gauge-independent result allows the commutation relations of the operators $\msgQ$ to be found.

Consider for example the elementary translation operators $\sgT_i$, which obey $\sgT_2 \sgT_1 = \sgT_1 \sgT_2$. To find the commutation relations of the corresponding MSG operators $\msgT_i$, consider $\pi_{\msgT_2^{-1}\msgT_1^{-1}\msgT_2\ns\msgT_1\ns}(j)$. The definition of $\pi_{\msgQ}$ gives the relation
\beq{EqpiMultiplication}
\pi_{\msgQ_2 \msgQ_1}(j) = \pi_{\msgQ_2}(\sgQ_1 j)\pi_{\msgQ_1}(j)\punc{,}
\eeq
which, together with \refeq{EqDefinepi2}, gives
\beq{EqscTexpand}
\pi_{\msgT_2^{-1}\msgT_1^{-1}\msgT_2\ns\msgT_1\ns}(j) = \omega(j \linkarrow \sgT_1 j \linkarrow \sgT_1 \sgT_2 j \linkarrow \sgT_2 j \linkarrow j)\punc{,}
\eeq
using the convention that, for second neighbors $i$ and $j$, $\link ij$ denotes the chain through their common neighbor. This expression involves a closed chain and so is gauge invariant; inspection of the example in \reffig{FigGaugeFieldDiagram} shows that it encloses a single diamond plaquette. This is in fact true for $j$ on either diamond sublattice, so one has $\pi_{\msgT_2^{-1}\msgT_1^{-1}\msgT_2\ns\msgT_1\ns}(j) = -1$ for any site $j$, independent of the gauge.

This is an instance of the general result that the MSG operator $\msgQ$ corresponding to a space-group operator $\sgQ$ is fixed by \refeq{EqDefinepi} only up to the arbitrary global phase of $\pi_{\msgQ}$. (As a special case, for any MSG operator $\msgE$ corresponding to the identity transformation $\sgE$, $\pi_{\msgE}(j)$ is equal to a uniform phase, independent of $j$.) If $-\msgQ$ denotes the MSG operator corresponding to the same space-group element as $\msgQ$ but with $\pi_{-\msgQ} = -\pi_{\msgQ}$, then one can write
\beq{EqTcommutation}
\msgT_2 \msgT_1 = -\msgT_1 \msgT_2\punc{.}
\eeq
One similarly finds $\msgT_1 \msgT_3 = -\msgT_3 \msgT_1$ and $\msgT_3 \msgT_2 = -\msgT_2 \msgT_3$. The commutation relations of these translations with other symmetry operations are calculated in \refapp{AppMSGCommutation}.

\subsubsection{Cubic translations}
\label{SecCubicTranslations}

Representations of the MSG are most easily found using an Abelian invariant subgroup of translation operators. The primitive FCC translations $\msgT_i$ do not commute with each other, so consider instead
\beq{EqDefineKi}
\msgK_i\ns = \msgT_i^{-2} \msgT_3\ns\msgT_2\ns\msgT_1\ns\punc{.}
\eeq
This combination ensures that $[\msgK_i,\msgT_j] = 0$ for all $i$ and $j$, and so $[\msgK_i,\msgK_j] = 0$. The corresponding space-group operators $\sgK_i$, which will be referred to as ``cubic translations'', translate by $4\uv_i$, the unit vectors of the simple cubic lattice upon which FCC is based. (The cubic unit cell, illustrated in \reffig{FigGaugeFieldDiagram}, contains four FCC sites and hence eight sites of the diamond lattice.)

It is shown in \refapp{AppMSGCommutation} that one can choose the global phases so that
\beq{EqphiK}
\pi_{\msgK_i}(j) \omega(\link{j}{\sgK_i j}) = 1\punc{,}
\eeq
where the chain $\link{j}{\sgK_i j}$ follows a right-handed helix; see \reffig{FigGaugeFieldDiagram} and \refeq{EqDefinephiK}. This allows the commutation relations of $\msgK_i$ with other any MSG operator $\msgQ$ to be found. Suppose some space-group operator $\sgQ$ obeys $\sgQ \sgK_i \sgQ^{-1} = \sgK'$; $\sgK'$ is then a cubic translation by $4\MQ \uv_i$. The commutation relation can be inferred from
\begin{widetext}
\begin{align}
\pi_{\msgK'^{-1} \msgQ \msgK_i \msgQ^{-1}}(j) &= \pi_{\msgK'}^*(j) \pi_{\msgQ}(\sgK_i \sgQ^{-1} j) \pi_{\msgK_i}(\sgQ^{-1} j) \pi^*_{\msgQ}(\sgQ^{-1} j)\\
&= \pi_{\msgK'}^*(j)\omega^*(j \linkarrow \cdots \linkarrow \sgK' j) \pi_{\msgK_i}(\sgQ^{-1} j) \omega(\sgQ^{-1} j \linkarrow \cdots \linkarrow \sgK_i \sgQ^{-1} j) \punc{.}\label{EqpiKQKQb}
\end{align}
\end{widetext}
By choosing a right-handed helix for the first chain, which spans a cubic translation from $j$ to $\sgK'j$, one can make use of \refeq{EqphiK}. For a proper transformation $\sgQ$, the second chain is also then a right-handed helix and one has $\pi_{\msgK'^{-1} \msgQ \msgK_i \msgQ^{-1}}(j) = 1$. By contrast, an improper transformation reverses chiralities, and inspection of the diamond lattice shows that the right- and left-handed spirals differ by their path around a single hexagonal plaquette. This leads finally to the important conclusion that
\beq{EqKQCommutation}
\msgQ \msgK_i = \msgK'\msgQ  \det \sgQ\punc{,}
\eeq
where $\det\sgQ=\det\MQ = \pm 1$.

\subsubsection{Time-reversal symmetry}
\label{SecTimeReversal1}

Besides the geometrical transformations of the pyrochlore (or diamond) lattice, another important symmetry of spin ice is that under reversal of all spins, which will be referred to as ``time reversal'' and denoted $\Theta$. As remarked above, this symmetry will be of profound importance for the eventual conclusions. In terms of the variables $B_\ell$, $\Theta$ obviously amounts to $B_\ell \leftarrow -B_\ell$ (for all diamond links $\ell$), which preserves the spin-ice constraint of \refeq{EqDivB} and is a symmetry of the configuration energy $\Ham_B$ provided $V$ contains only even powers of $B$. (This will not be the case if, for example, an external magnetic field is applied to the spins.)

\refeq{EqBfromA} gives, correspondingly, $A_\lambda \leftarrow -A_\lambda$, under which the last two terms of $\Ham_{A,\theta}$, in \refeq{EqEA2}, are again trivially symmetric (for appropriate $V$). Considering once more $\Ham^{(\Lambda)}$ in \refeq{EqEA3}, it is necessary to transform $\psi_j$ according to
\beq{EqTRpsi}
\psi_j \xleftarrow{\Theta} \psi_j^* \pi_\Theta(j)
\punc{,}
\eeq
so $\Theta$ is an antiunitary operator. Time reversal is its own inverse, so $\Theta^2 = 1$, implying that $\lvert\pi_\Theta(j)\rvert = 1$. The phase should be chosen such that
\beq{EqDefinepiTheta}
\pi_{\Theta}\ns(i)\pi_\Theta^*(j) \omega^*(\link ij) = \omega^*(\link ji)\punc{,}
\eeq
and the terms in $H^{(\Lambda)}$ for $\link ij$ and $\link ji$ are exchanged by $\Theta$. Note that if the background gauge potential is chosen real, then $\omega(\link i j) = \omega(\link ji)$ and one can set $\pi_\Theta = 1$.

Applying $\Theta$ and the MSG operator $\msgQ$ successively to $\psi_j$ shows that the two commute only if
\beq{EqThetaQcommutation}
\pi\ns_{\msgQ}(\sgQ^{-1}j)\pi\ns_\Theta(\sgQ^{-1}j) = \pi\ns_\Theta(j)\pi^*_{\msgQ}(\sgQ^{-1}j)\punc{,}
\eeq
or equivalently, using \refeq{EqDefinepiTheta}, if
\beq{EqThetaQcommutation2}
[\pi_{\msgQ}(j)\omega(\link{j}{\sgQ j})]^2 = 1\punc{,}
\eeq
for all sites $j$. In fact, it is straightforward to show, using \refeq{EqDefinepi2} and the fact that $\omega(i \linkarrow \cdots \linkarrow i) = \pm 1$, that if this is true for any site, it is true for all sites. By choosing the global phase of $\pi_{\msgQ}$ so that $\pi_{\msgQ}(0)\omega(\link{0}{\sgQ 0})$ is real, one can therefore ensure that $\Theta$ commutes with any MSG operator $\msgQ$. This condition is indeed satisfied by the conventions used here for the translation operators $\msgT_i$ and $\msgK_i$, according to \refeqand{EqphiTij}{EqphiK2b}.

\section{MSG representations}
\label{SecRepresentations}

The Higgs transitions involve condensation of the matter field $\psi_j$, and the nature of the ordered phase is determined by the configuration of the condensate. While the details will depend on microscopics, the symmetry structure can be elucidated by considering the irreducible representations, or ``irreps'', of the MSG, which label the modes of the matter field. The irreps will be found by following the general procedure for space groups,\cite{Cornwell} but there will be certain differences because of the commutation relations such as \refeq{EqTcommutation}.\cite{EndNoteDoubleGroup} No attempt will be made to find all representations systematically; interest will be restricted to those with smallest dimension for certain high-symmetry wavevectors.

As remarked in \refsec{SecCubicTranslations}, it is convenient to construct representations starting from an invariant subgroup of mutually commuting translation operators, such as the cubic translations $\msgK_i$. The vectors belonging to a given representation will be labeled by their eigenvalue of the operators $\msgK_i$, according to
\beq{EqDefineKetkappa}
\msgK_i \ket{\kappav} = \ee^{-\ii \kappav \cdot (4\uv_i)}\ket{\kappav}\punc{;}
\eeq
note that $\sgK_i$ translates by $4\uv_i$. Distinct representations are associated with ``wavevectors'' $\kappav$ within the region where $-\frac{\pi}{4} \le \kappav \cdot \uv_i < \frac{\pi}{4}$ for all $i$. This cubic ``reduced Brillouin zone'' $\BZR$ has a volume of $1/4$ that of the Brillouin zone $\BZL$ of the FCC Bravais lattice (a truncated octahedron). The notation $\reducedR{\kappav}$ will be used to signify wavevector $\kappav$ reduced to $\BZR$ by addition of multiples of the reciprocal lattice vectors $\frac{\pi}{2}\uv_i$.

\subsection{FCC translations}

The FCC translations $\msgT_i$ commute with $\msgK_i$, so one can also label the vectors by their eigenvalue under $\msgT_3$, say. From \refeq{EqDefineKi}, one has $\msgT_3^2 = -\msgK_1 \msgK_2$, so for given $\kappav$ there are only two possible eigenvalues of $\msgT_3$,
\beq{EqDefineKetkappaell}
\msgT_3 \ket{\kappav, \ell} = \ii (-1)^{\ell} \ee^{-\ii \kappav \cdot \ev_3}\ket{\kappav,\ell}\punc{,}
\eeq
where $\ell \in \{0,1\}$.

Since the remaining FCC translation operators $\msgT_1$ and $\msgT_2$ commute with $\msgK_i$ but not $\msgT_3$, they mix vectors with equal $\kappav$ but distinct $\ell$. For instance, $\msgT_3 \msgT_1 = -\msgT_1 \msgT_3$, so the vector $\msgT_1 \ket{\kappa, \ell}$ has $\msgT_3$-eigenvalue $-\ii(-1)^{\ell}\ee^{-\ii \kappav\cdot\ev_3}$, allowing one to define
\beq{EqRelateTwoells}
\ket{\kappav,1} = -\ii \ee^{-\ii \kappav \cdot \ev_1} \msgT_1\ket{\kappav,0}\punc{,}
\eeq
which implies
\beq{EqRelateTwoells2}
\ket{\kappav,0} = -\ii \ee^{-\ii \kappav \cdot \ev_1} \msgT_1\ket{\kappav,1}\punc{.}
\eeq
The action of $\msgT_2$ on the eigenvectors can then be found by using $\msgT_2\ns = -\msgK_3\ns\msgT_1^{-1}\msgT_3\ns$. These results can be summarized as
\beq{EqTketkappaell}
\msgT_i \ket{\kappav, \ell} = \ii \sum_{\ell'} \sigma^i_{\ell'\ell} \ee^{-\ii \kappav \cdot \ev_i}\ket{\kappav,\ell'}\punc{,}
\eeq
where $\sigmam^{1,2,3}$ are the Pauli matrices.

\subsection{Time reversal}
\label{SecTimeReversal2}

The time-reversal operator $\Theta$, introduced in \refsec{SecTimeReversal1}, is antiunitary and commutes with the translation operators $\msgT_i$ and $\msgK_i$. Applying $\Theta$ to \refeq{EqDefineKetkappa} one therefore finds that $\Theta \ket{\kappav}$ is also an eigenvector of $\msgK_i$, but with wavevector $\reducedR{-\kappav}$. (Reduction to $\BZR$ is only strictly necessary for $\kappav$ at the edge of $\BZR$, but these points will be of particular importance.)

One can therefore define the matrix $\thetam(\kappav)$,
\beq{EqDefthetam}
\Theta \ket{\kappav,\ell} = \sum_{\ell'} \theta_{\ell'\ell}(\kappav) \ket{\reducedR{-\kappav},\ell'}\punc{.}
\eeq
The commutation of $\Theta$ and $\msgT_i$ then gives
\beq{EqthetaTicommutation}
\sigmam^i \thetam(\kappav) \ee^{-\ii \reducedR{-\kappav}\cdot \ev_i} = -\thetam(\kappav) (\sigmam^i)^* \ee^{\ii \kappav \cdot \ev_i}\punc{,}
\eeq
for all $i \in \{1,2,3\}$. This is sufficient to fix $\thetam(\kappav)$ up to phase, but $\thetam(\kappav)$ in fact depends on the relation between $[-\kappav]$ and $\kappav$, and can be written as
\beq{Eqthetam}
\thetam(\kappav) = \vartheta(\kappav) \times
\begin{cases}
\sigmam^2&\text{if $\reducedR{-\kappav} = -\kappav$ (body of $\BZR$)}\\
\sigmam^j\sigmam^2&\text{if $\reducedR{-\kappav} = -\kappav - \frac{\pi}{2}\uv_j$ (face)}\\
\sigmam^j\sigmam^2&\text{if $\reducedR{-\kappav} = -\kappav - \frac{\pi}{4}\ev_j$ (edge)}\\
\sigmam^2&\text{if $\reducedR{-\kappav} = -\kappav - \frac{\pi}{2}\deltav_0$ (corner),}\\
\end{cases}
\eeq
where $\lvert\vartheta(\kappav)\rvert = 1$.

For consistency, one requires $\Theta^2 = 1$, which relates the values of $\vartheta(\kappav)$ and $\vartheta(-\kappav)$ at a generic wavevector $\kappav$. For the points $\kappav = \zerov$ and $\kappav = -\frac{\pi}{4}\deltav_0$ (the center and corner of $\BZR$), however, \refeq{Eqthetam} gives a contradiction. This implies that there must be an enlarged (Kramers) degeneracy at these points, and another label besides $\ell$ is required. The hopping model studied in \refapp{AppHoppingModel} has Dirac cones in the spectrum around these points, confirming this result.

\subsection{Other transformations}
\label{SecRepsOthers}

The action of the remaining MSG operators $\msgQ$ on a vector $\ket{\kappav,\ell}$ can be determined using the commutation relations of $\msgQ$ with $\msgT_i$ and $\msgK_i$. Treating first the $\kappav$ label, acting with $\msgQ$ on the defining equation \refeq{EqDefineKetkappa} and using \refeq{EqKQCommutation} gives
\beq{EqQKket}
\msgK' \msgQ \ket{\kappav,\ell}\det \sgQ = \ee^{-\ii \kappav \cdot (4\uv_i)} \msgQ \ket{\kappav,\ell}\punc{.}
\eeq
The operator $\msgK'$ translates by $4\MQ \uv_i$, so $\msgQ \ket{\kappav,\ell}$ has wavevector $\kappav_{\sgQ}(\kappav)$, where $\ee^{-\ii \kappav_{\sgQ}(\kappav)\cdot(4\MQ \uv_i)} \det \sgQ = \ee^{-\ii \kappav\cdot(4\uv_i)}$. For $\det \sgQ = 1$, the wavevector therefore transforms as usual under the space-group operation $\sgQ$, with $\kappav_{\sgQ}(\kappav) = \reducedR{\MQ \kappav}$. For improper transformations, the factor of $\det\sgQ=-1$ appearing in the commutation relation leads to $\kappav_{\sgQ}(\kappav) = \reducedR{\MQ \kappav - \frac{\pi}{4}\deltav_0}$.

The set of distinct wavevectors given by $\kappav_{\sgQ}(\kappav)$ for all $\sgQ$ is referred to as $\star\kappav$, the ``star'' of $\kappav$. Labeling the wavevectors within a star by $\kappav_{\kappa}$ for $\kappa \in \{ 1, 2, \ldots, \lvert\star\kappav\rvert\}$, one can write
\beq{EqDefineLambda}
\msgQ \ket{\kappav_{\kappa}, \ell} = \sum_{\kappa'\ell'} \Lambda_{\kappa'\ell',\kappa\ell}(\msgQ) \ket{\kappav_{\kappa'},\ell'}\punc{,}
\eeq
defining the matrix $\Lambdam$, with $2\lvert\star \kappav\rvert$ rows and columns. Applying a second MSG operator $\msgQ'$ to \refeq{EqDefineLambda} gives
\beq{EqLambdaProduct}
\Lambdam(\msgQ'\msgQ) = \Lambdam(\msgQ')\Lambdam(\msgQ)\punc{,}
\eeq
so these matrices provide a unitary representation of the MSG. (This is also therefore a projective representation of the space group, with exactly the same factor system as the MSG.) According to \refeqand{EqDefineKetkappa}{EqTketkappaell}, one has $\Lambda_{\kappa'\ell',\kappa\ell}(\msgK_i) = \ee^{-\ii\kappav_\kappa\cdot (4\uv_i)}\delta_{\kappa\kappa'}\delta_{\ell\ell'}$ and
\beq{EqLambdaTi}
\Lambda_{\kappa'\ell',\kappa\ell}(\msgT_i) = \ii \ee^{-\ii\kappav_\kappa\cdot \ev_i} \delta_{\kappa\kappa'} \sigma^i_{\ell'\ell}\punc{;}
\eeq
it is similarly useful to define $\Lambdam(\Theta)$.

The matrix $\Lambdam(\msgQ)$ is in principle determined by the commutation relations of $\msgQ$ with $\msgT_i$, but this calculation is complicated by the need to make a consistent assignment of the relative phase of $\ket{\kappav,\ell}$ between different wavevectors $\kappav_\kappa$. It is simpler to use the transformation properties of the eigenstates of the hopping model treated in \refapp{AppHoppingModel}, which has the same symmetry group and so the same set of representations. The main interest, however, is not the transformation properties of the vectors $\ket{\kappav,\ell}$, but rather those of the gauge-invariant combinations of these, to be introduced in \refsec{SecBilinears}. These can in certain cases (the diagonal bilinears) be calculated using only the commutation relations of the matrices $\Lambdam$, which are identical to those of the MSG operators $\msgQ$.

\subsection{Relevant matter-field modes}
\label{SecMatterFieldModes}

The stars $\star\kappav$ play an important role in the long-wavelength description: if a mode of the matter field with wavevector $\kappav$ is relevant at a transition, then symmetry requires that the corresponding mode at $\kappav_{\sgQ}(\kappav)$ is also. The set of matter fields appearing in the action will therefore generically consist of a single star $\star\kappav$. If the symmetry group is reduced by the perturbations $V$ in \refeq{EqSpinIceHamiltonian}, then the number of wavevectors in a star can be correspondingly reduced (see \refsec{SecSingleWavevector}).

For a generic point, $\star\kappav$ has $96$ members (the point group has $48$ elements; time reversal contributes a factor of $2$), but certain high-symmetry points have considerably smaller stars. The points $\kappav = \zerov$ and $-\frac{\pi}{4}\deltav_0$ are linked by improper transformations and form a star consisting of only two wavevectors. As noted in \refsec{SecTimeReversal2}, however, time-reversal symmetry increases the dimensionality of the corresponding representations and the presence of Dirac cones at these points in the hopping model studied in \refapp{AppHoppingModel} makes Higgs transitions involving these modes appear unlikely.

The only other wavevectors that are invariant under time reversal are $-\frac{\pi}{4}\uv_i$ and $-\frac{\pi}{8}\ev_i$, $i \in \{1,2,3\}$, which form separate stars with respect to the space group but are linked by improper MSG transformations. (The two sets are equivalent and can be exchanged by choosing the opposite sign in \refeq{EqDefineKi}.) They therefore form a star with $6$ members, to be denoted $\star\kappav_6$; all other stars consist of more than $6$ wavevectors.

As described in detail in \refsec{SecOrderParameters}, the symmetry of an ordered Higgs phase depends on that of the wavevector(s) $\kappav$ at which condensation occurs. Of most interest in spin ice are those ordered states with definite sign under time reversal (odd for spin-ordered, even for paramagnetic); such states require that condensation occurs at time-reversal-symmetric wavevectors. The concrete results presented here are therefore based on the assumption that $\star\kappav_6$ is the unique star that is relevant at the transition. (These modes indeed have minimal energy for a range of parameters in the hopping model of \refapp{AppHoppingModel}.) Exactly the same procedure can be applied when other stars are relevant.

It is convenient to introduce an alternative labeling for the wavevectors in $\star\kappav_6$: let $\kappav_{j+} = -\frac{\pi}{4}\uv_j$ and $\kappav_{j-} = -\frac{\pi}{8}\ev_j$ for $j \in \{1,2,3\}$. The reason for this notation is that, as described in \refsec{SecOrderParameters}, the matter fields at these six wavevectors can be used to construct order parameters for six spin spirals (see \reffig{FigSpinSpiral}), parallel to the three cubic axes and with right- and left-handed chiralities.

\subsection{Continuum fields}
\label{SecContinuumFields}

The physics of the Coulomb phase is described by a long-wavelength theory found by coarse-graining\cite{Youngblood,Huse,Henley,Isakov1} the gauge field $A_\lambda$. A theory of the Higgs transitions must also involve the matter fields $\psi_j$, whose mode structure carries information about the microscopic details that is important in the ordered phase.

The set of matter-field modes that are relevant to the Higgs transition form a representation of the MSG, so the notation $\ket{\kappav_\kappa, \ell}$ will henceforth be used to denote such a mode. (In the generic case, no further indices are required to label the complete set of relevant modes.) The important configurations of the matter field are those that are ``close to'' these modes, and can be written using the same notation as
\beq{EqDefinevarphi}
\ket{\psi} = \int{\dd^3\rv} \sum_{\kappa,\ell} \varphi_{\kappa\ell}(\rv) \Proj_{\rv} \ket{\kappav_{\kappa},\ell}\punc{,}
\eeq
where $\Proj_{\rv}$ is a projector that restricts to a large region surrounding the point $\rv$. This expression defines the continuum field $\phiv(\rv)$, a $2\lvert\star \kappav\rvert$-component vector, which is assumed to be slowly varying on the lattice scale. Under a spatial symmetry $\sgQ$, the configuration is transformed to $\msgQ\ket{\psi}$, giving the corresponding transformation
\beq{EqTransformphiv}
\phiv(\rv) \xleftarrow{\msgQ} \Lambdam(\msgQ) \phiv(\sgQ^{-1}\rv)
\eeq
for the continuum field.

The continuum version of the gauge field $A_\lambda$ will be denoted $\alphav(\rv)$ and transforms as a vector,
\beq{Eqalphavtransformation}
\alphav(\rv) \xleftarrow{\sgQ} \MQ \alphav(\sgQ^{-1} \rv)\punc{,}
\eeq
under the space-group operation $\sgQ$. The microscopic gauge symmetry \refeq{EqGaugeSymmetry2} then transforms the continuum fields $\alphav$ and $\phiv$ as
\beq{EqGaugeContinuum}
\begin{aligned}
\alphav &\leftarrow \alphav + \parv X(\rv)\\
\phiv &\leftarrow \phiv\ee^{\ii X(\rv)}
\end{aligned}
\eeq
(where a factor of $2\pi$ has been eliminated by the definition of $\alphav$). It is therefore convenient to introduce the ``covariant derivative''
\beq{EqCovariantDerivative}
\Dv = \parv - \ii \alphav\punc{,}
\eeq
so $D_\mu \varphi_{\kappa\ell}$ is gauge invariant (but transforms like $\phiv$ under global phase rotations). The operator $\Dv$ is also a vector under the space group.

Under time reversal, the matter-field configuration $\ket{\psi}$ is replaced by $\Theta \ket{\psi}$, so the continuum field transforms as $\phiv \xleftarrow{\Theta} \Lambdam(\Theta)\phiv^*$. As noted in \refsec{SecTimeReversal1}, $A_\lambda$ is odd under time reversal; the same is true for $\alphav$, so
\beq{EqDtimereversal}
D_\mu \phiv \xleftarrow{\Theta} \Lambdam(\Theta)(D_\mu \phiv)^*\punc{.}
\eeq

\section{Order parameters}
\label{SecOrderParameters}

The continuum fields $\phiv(\rv)$ and $\alphav(\rv)$ provide the desired long-wavelength description, but the strongly coupled nature of the system prevents a direct derivation of the relation between these and physical observables. Each observable can instead be associated with a function of the continuum fields with identical symmetries, and the two are, to leading order, proportional.

It might appear most natural to express the order parameters in terms of the gauge field $A$, related to the spins $\Sv(\rv)$ by \refeq{EqBfromA}, or its continuum equivalent $\alphav$. The Higgs transition is, however, more readily described in terms of the matter fields $\phiv$, and it will be shown that the order parameters can also be expressed as combinations of these. Because of the gauge symmetry under \refeq{EqGaugeContinuum}, the lowest-order combinations that can correspond to physical quantities are bilinears of $\phiv$.

\subsection{Gauge-invariant bilinears}
\label{SecBilinears}

Given the set of $2\lvert\star \kappav\rvert$ matter fields $\varphi_{\kappa\ell}(\rv)$, one can construct all bilinears as
\beq{EqDefinePhi}
\Phi_{\mu \kappa \kappa'}(\rv) = \phiv_{\kappa}^\dagger(\rv) \sigmam^{\mu} \phiv\ns_{\kappa'}(\rv)\punc{,}
\eeq
where $\mu \in \{0,\ldots,3\}$ and $\sigmam^0$ is the $2\times 2$ identity. (In this expression, the matrix notation applies only to the $\ell$ indices.) These quantities are not independent, satisfying $\Phi_{\mu\kappa\kappa'}^* = \Phi_{\mu\kappa'\kappa}\ns$ and
\beq{EqPhiRedundancy}
\sum_{\mu = 0}^3 \Phi_{\mu\kappa_1\ns\kappa_1'}^* \Phi\ns_{\mu\kappa_2\ns\kappa_2'} = 2\Phi_{0\kappa_1\kappa_2}^*\Phi_{0\kappa_1'\kappa_2'}\ns\punc{.}
\eeq

From \refeq{EqTransformphiv}, the bilinears transform according to
\beq{EqTransformPhi}
\Phiv(\rv) \xleftarrow{\sgQ} \Deltam(\sgQ) \Phiv(\sgQ^{-1}\rv)\punc{,}
\eeq
where the matrix $\Deltam$ is given by
\beq{EqDefineDelta}
\Delta_{\mu\kappa_1\ns\kappa_2\ns,\nu\kappa_1'\kappa_2'}(\sgQ) = \frac{1}{2} \Tr \Lambdam^\dagger_{\kappa_1\ns\kappa_1'}(\msgQ)\sigmam^\mu \Lambdam\ns_{\kappa_2\ns\kappa_2'}(\msgQ)\sigmam^\nu\punc{.}
\eeq
Note that the overall phase of the MSG operator $\msgQ$ has no effect on $\Deltam(\sgQ)$, and that these matrices form a unitary (vector, rather than projective) representation of the space group.

One can therefore associate with each bilinear a wavevector referred to the full FCC Brillouin zone $\BZL$. Using \refeq{EqLambdaTi}, one finds
\beq{EqDeltaTi}
\Delta_{\mu\kappa_1\ns\kappa_2\ns,\nu\kappa_1'\kappa_2'}(\sgT_i) = \delta_{\mu\nu}\delta_{\kappa_1\ns\kappa_1'}\delta_{\kappa_2\ns\kappa_2'} \ee^{-\ii \ev_i \cdot (\kappav_{\kappa_2} - \kappav_{\kappa_1} - \frac{\pi}{2}\uv_\mu)}\punc{,}
\eeq
so the bilinear $\Phi_{\mu\kappa_1\kappa_2}$ has wavevector $\kappav_{\kappa_2} - \kappav_{\kappa_1} - \frac{\pi}{2}\uv_{\mu}$ (where $\uv_0 = \zerov$). For a given star $\star \kappav$, the $\BZL$-wavevectors appearing in the representation provided by $\Deltam(\sgQ)$ are given by all such combinations of the $\BZR$-wavevectors in $\star\kappav$. These will comprise more than one star under $\sgQ$ (i.e., more than one closed set under the action of the matrices $\MQ$) and so the representation $\Deltam(\sgQ)$ is necessarily reducible, even when $\Lambdam(\msgQ)$ is irreducible.

Of particular interest are the ``diagonal'' bilinears of the form $\Phi_{\mu\kappa\kappa}$, where the two wavevectors are identical. These are real and obey $\sum_{\mu = 1}^3 \Phi_{\mu\kappa\kappa}^2 = \Phi_{0\kappa\kappa}^2$, allowing each bilinear to be treated as a $3$-component vector $\Phiv_\kappa$, with magnitude given by $\lvert\Phiv_{\kappa}\rvert = \Phi_{0\kappa\kappa}$. The wavevector associated with $\Phi_{\mu\kappa\kappa}$ is $-\frac{\pi}{2}\uv_\mu$, irrespective of $\kappa$, and these are clearly the only bilinears whose wavevector vanishes on reduction to $\BZR$. The diagonal bilinears therefore correspond to order parameters whose unit cell is no larger than the cubic unit cell of the pyrochlore lattice (containing $16$ sites).

The wavevectors $-\frac{\pi}{2}\uv_\mu$ are closed under the action of the space-group symmetries, so the transformation matrices $\Delta_{\mu\kappa\kappa,\nu\kappa'\kappa'}$ for this subset form a representation of the space group (a ``subrepresentation'' of that provided by $\Deltam$). As noted in \refsec{SecRepsOthers}, these matrix elements can be calculated directly based on the multiplication table of $\Lambdam(\msgQ)$. From \refeq{EqDefineDelta},
\begin{multline}
\label{EqDeltaDiagonal}
\Delta_{\mu\kappa\kappa,\nu\kappa'\kappa'}(\sgQ) = -\frac{\ii^{\delta_{\mu0}+\delta_{\nu0}}}{2} \ee^{\ii(\kappav_\kappa\cdot\ev_\mu+\kappav_{\kappa'}\cdot\ev_\nu)}\\{}\times\Tr\Lambdam^\dagger_{\kappa\kappa'}(\msgQ)\Lambdam\ns_{\kappa\kappa}(\msgT_\mu)\Lambdam\ns_{\kappa\kappa'}(\msgQ)\Lambdam\ns_{\kappa'\kappa'}(\msgT_\nu)\punc{,}
\end{multline}
where the Pauli matrices have been expressed in terms of $\Lambdam(\msgT_i)$ using \refeq{EqLambdaTi}. The matrices $\Lambdam$ form a unitary representation of the MSG, so this can be rewritten as
\begin{multline}
\label{EqDeltaDiagonal2}
\Delta_{\mu\kappa\kappa,\nu\kappa'\kappa'}(\sgQ) = -\frac{\ii^{\delta_{\mu0}+\delta_{\nu0}}}{2} \ee^{\ii(\kappav_\kappa\cdot\ev_\mu+\kappav_{\kappa'}\cdot\ev_\nu)}\\{}\times
\delta_{\kappav_\kappa,\kappav_\sgQ(\kappav_{\kappa'})}
\Tr\Lambdam_{\kappa'\kappa'}(\msgQ^{-1}\msgT_\mu\msgQ\msgT_\nu)\punc{.}
\end{multline}
The operator $\msgQ^{-1}\msgT_\mu \msgQ \msgT_\nu$ gives a pure translation by $\ev_\nu + \MQ\ev_\mu$, and can be simplified using \refeq{EqpiTQTQ3}. The matrix element $\Delta_{\mu\kappa\kappa,\nu\kappa'\kappa'}(\sgQ)$ is finally found by once more applying \refeq{EqLambdaTi}.

The transformation properties of $\Phiv(\rv)$ under time reversal can be found straightforwardly using the results of \refsec{SecTimeReversal2}. For the diagonal bilinears involving $\star\kappav_6$, for example, one has
\beq{EqPhiTheta}
\Phi_{\mu,j\pm,j\pm} \xleftarrow{\Theta} (-1)^{\delta_{\mu j}}\Phi_{\mu,j\pm,j\pm}\punc{,}
\eeq
where the notation introduced in \refsec{SecMatterFieldModes} is used. There is therefore a single diagonal bilinear that is odd under time reversal at each wavevector $\kappav_{j\pm}$.

\subsection{Observables}
\label{SecObservables}

To identify the bilinears discussed above with appropriate physical quantities, it is necessary to find an observable that transforms under the symmetries of the problem in the same way as a given bilinear. The first step is therefore to identify the potential order parameters in the physical model and determine their transformation properties.

The observables of primary interest are of course the spin degrees of freedom $\Sv_{\rv}$. These are odd under time reversal (by definition of the latter) and are pseudovectors: under a space-group symmetry $\sgQ$,
\beq{EqSpinTransformation}
\Sv_{\rv} \xleftarrow{\sgQ} (\det \sgQ) \MQ \Sv_{\sgQ^{-1}\rv}\punc{.}
\eeq
One can alternatively make the Ising nature of the spins explicit using the scalar variable $s_{\rv}$ defined in \refsec{SecSpinIce}, which transforms as $s_{\rv} \leftarrow (\det \sgQ)(-1)^{\sigmaQ}s_{\sgQ^{-1}\rv}$.

Those bilinears that are even under time reversal cannot correspond to spin degrees of freedom, so one is led to consider also correlations of the spins. (Such observables can be order parameters for paramagnetic ordered phases, as discussed in \refsec{SecOrderedStates}.) In particular, the product of two nearest-neighbor spins $\Sv_{\rv_1}\cdot \Sv_{\rv_2} = \pm \frac{1}{3}$ is time-reversal even and transforms as a scalar under the space group. Within the ice-rule manifold, every tetrahedron has two edges where this is negative, on opposite sides of the tetrahedron. (These are the frustrated ferromagnetic bonds, linking sites with the same value of the Ising variable $s$.) One can therefore combine these nearest-neighbor correlations into a nematic variable $\Nv_i$ for tetrahedron $i$, which lies along the cubic direction joining the edges with negative correlation and transforms as $\Nv_i \leftarrow \pm \MQ \Nv_{\sgQ i}$.

\begin{figure}
\putinscaledfigure{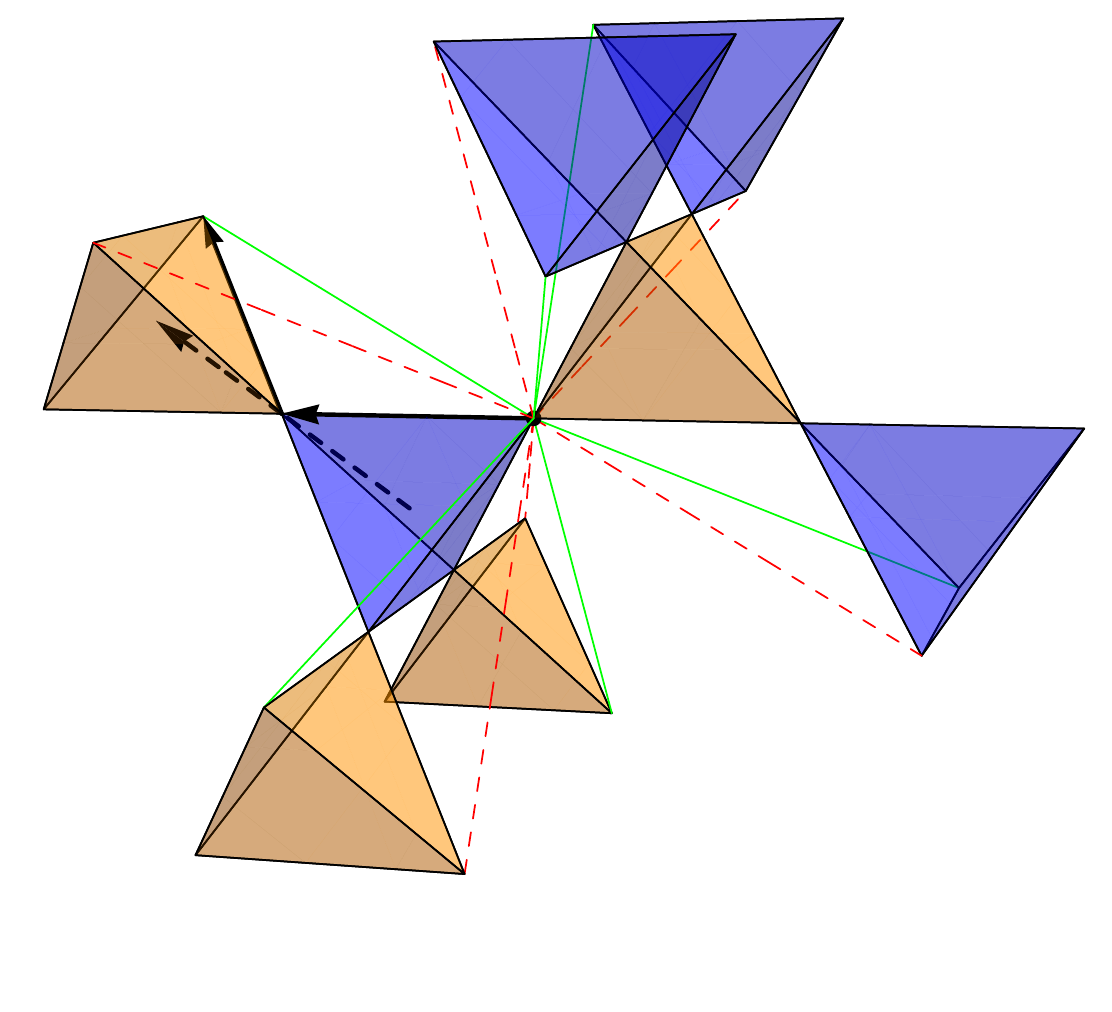}
\caption{The $12$ second-nearest neighbors of a site of the pyrochlore lattice. The second-neighbor pairs can be separated into two ``pinwheels'' of opposite chirality, joined with solid (green) and dashed (red) lines. (The chirality is defined by the scalar triple product of the three vectors illustrated with arrows: the two solid arrows link a pair via their common neighbor and the dashed arrow links the centers of their tetrahedra.) The difference between the spin correlations on the two pinwheels gives a pseudoscalar order parameter, labeled ``$-$'' in \reftab{TabOrderParameters}.
\label{FigPinwheel}}
\end{figure}

As illustrated in \reffig{FigPinwheel}, each site of the pyrochlore lattice has $12$ second-nearest neighbors, separated by displacements of $\frac{1}{2}(\ev_1 + \ev_2)$ and equivalent vectors. General combinations of correlations between second-neighbor spins will not be treated, but it is worth observing the pairs can be divided into two ``pinwheels'' of opposite chirality. While the symmetric combination of all correlations is clearly a scalar quantity, the difference between correlations on right- and left-handed pinwheels is a pseudoscalar (changing sign under improper transformations).

Further-neighbor pairs of spins and higher-order correlations can also be considered, but these are subject to rapidly diminishing returns. One is also limited by the restriction to the ice-rule manifold; if this were relaxed, the monopole density would, for example, provide a scalar order parameter.

\subsection{Irreducible representations}
\label{SecIrreps}

The observables can be combined into a vector $\Fv$, so that any linear combination can be expressed as $\vv \cdot \Fv$, where the vector $\vv$ transforms according to matrices $\Dm(\sgQ)$ determined by the considerations of the previous section. One therefore has two representations of the space group, given by the transformation matrices for the bilinears $\Deltam(\sgQ)$ and those for the observables. These two sets of quantities will be identified by expressing both in terms of the irreps of the space group.\cite{Cornwell}

An irrep $\Gammam^{\kv p}$ is labeled by a $\BZL$-wavevector $\kv$ and an index $p$, and consists of a set of $d_{\kv p}\times d_{\kv p}$ matrices $\Gammam^{\kv p}(\sgQ)$ that obey the multiplication table of the space group. Since the translation operators $\sgT_i$ do not commute with the other operators $\sgQ$, a given irrep involves all wavevectors in the star $\star \kv$. (The symbol $\kv$ will be used for wavevectors referred to $\BZL$ and $\star \kv$ for the set of wavevectors resulting from the action of $\MQ$ on $\kv$.) The matrices for translation operators $\Gammam^{\kv p}(\sgT)$ are diagonal, with the element in row $a$ given by $\ee^{-\ii \kv_a}$, where $\kv_a \in \star \kv$.

The projection matrices given by\cite{Cornwell}
\beq{EqProjD}
\ProjD{}^{\kv p}_{ab} = \frac{d_{\kv p}}{\lvert\spacegroup\rvert} \sum_{\sgQ\in\spacegroup} [\Gamma^{\kv p}_{ab}(\sgQ)]^* \Dm(\sgQ)
\eeq
can be used to find a set of observables that transform according the irrep $\Gammam^{\kv p}$. (The equivalent set of matrices for the bilinears will be denoted $\ProjDelta{}^{\kv p}_{a b}$.) In particular, for any choice of $\vv$, the vectors $\vv_a = \ProjD{}^{\kv p}_{a1} \ProjD{}^{\kv p}_{11} \vv$ for $a \in \{1,2,\ldots,d_{\kv p}\}$ form a set that transform under $\sgQ$ according to the matrix $\Gammam^{\kv p}(\sgQ)$:
\beq{EqFaTransformation}
\Dm(\sgQ) \vv_a = \sum_b \Gamma^{\kv p}_{ba}(\sgQ) \vv_b\punc{.}
\eeq
The observable $F^{\kv p}_a = \vv_a \cdot \Fv$ is then said to transform as row $a$ of the representation $\Gammam^{\kv p}$. (While this statement is formally true for any choice of starting vector $\vv$, one can certainly choose it such that $\vv_a = \zerov$ for all $a$, in which case the result is of course trivial. If there are different starting vectors $\vv$ and $\vv'$ that give linearly independent results $\vv_a\ns$ and $\vv_a'$, these are said to provide multiple ``copies'' of the representation.) The corresponding combination of the bilinears, denoted $\Phi^{\kv p}_a$, transforms in exactly the same way as $F^{\kv p}_a$, allowing the pair to be identified up to a constant of proportionality.

Time-reversal symmetry can be included by doubling the number of representations, with each labeled either even or odd under $\Theta$. (The full group is a direct product of the time-reversal group and the space group; the former is Abelian.) Each projection operator will then include a factor projecting into the subspace with appropriate sign under time reversal.

Including this factor and rewriting the sum in \refeq{EqProjD} in terms of the subgroup of translations $\Tgroup$ and the coset representatives $\Qgroup$ (defined in \refsec{SecLatticeSymmetries}) gives
\begin{multline}
\label{EqProjD2}
\ProjD{}^{\kv p {\pm}}_{ab} = \frac{1}{2}\left[\Dm(\sgE) \pm \Dm(\Theta)\right]\left[\frac{1}{\lvert\Tgroup\rvert} \sum_{\rv} \ee^{\ii \kv_a \cdot \rv}\Dm(\sgT_{\rv})\right]\\{}\times
\left[\frac{d_{\kv p}}{\lvert\Qgroup\rvert}\sum_{\sgQ\in\Qgroup} [\Gamma^{\kv p}_{ab}(\sgQ)]^* \Dm(\sgQ)\right]\punc{,}
\end{multline}
where $\pm$ denotes an irrep that is even or odd under time reversal. The second factor in this product involves an infinite sum over translations $\sgT_{\rv}$ by all FCC-lattice vectors $\rv$, and has the effect of projecting into the subspace of observables with wavevector $\kv_a$.

There are $24$ diagonal bilinears that can be constructed using the wavevectors comprising $\star\kappav_6$, defined in \refsec{SecMatterFieldModes}, and the linear combination for each irrep is found using the analogue of \refeq{EqProjD2}. Results are presented in \reftab{TabOrderParameters} for the diagonal bilinears, which, as noted in \refsec{SecBilinears}, have wavevectors $\kv = -\frac{\pi}{2}\uv_\mu$ for $\mu \in \{0,\ldots,3\}$. These can be divided into two stars under the space group, to be denoted $\star\kv_{\spG} = \{\zerov\}$ and $\star\kv_{\spX} = \{-\frac{\pi}{2}\uv_i\}$, corresponding to the symmetry points $\spG$ and $\spX$ respectively. Observables with these wavevectors have period no larger than the cubic unit cell of the pyrochlore lattice and will provide the order parameters for the phases of interest in \refsec{SecPhaseTransitions}.
\begin{table*}
\caption{Order parameters for transitions in spin ice and the corresponding diagonal bilinears constructed from the modes of the matter field at wavevectors $\star\kappav_6 = \{\kappav_{j\chi}\}$ where $j \in \{1,2,3\}$ and $\chi = \pm 1$. The irreps of $O_h^7$ are labeled using the standard Bouckaert convention,\cite{Cornwell} with the superscript specifying the sign under time reversal. The nematic observables, illustrated in \reffig{FigdWave}, are labeled $d_{j\tau}$ and $\bar{d}_{j\tau}$, where $\tau$ denotes the structure on the two tetrahedron orientations (correlations on tetrahedra of a single orientation for $\tau = 0,1$; symmetric or antisymmetric combinations for $\tau = \pm$). The vectors $\vv$ are defined by $\vv(d_1) = \frac{1}{\sqrt{3}}(1,{-\frac{1}{2}},{-\frac{1}{2}})$, $\vv(\bar{d}_1) = \frac{1}{2}(0,1,{-1})$. All sums are over $j \in \{1,2,3\}$ and $\chi = \pm$; the diagonal bilinear $\Phi_{i,j\chi,j\chi}$ is denoted $\Phi_{i,j\chi}$ for brevity.}
\label{TabOrderParameters}
\begin{ruledtabular}
\begin{tabular}{cccl}
irrep $\Gamma$ & observables $F^{\Gamma}_a$ & bilinears $\Phi^{\Gamma}_a$ & description\\\hline
$\spG_1^+$ & $+$ & $\frac{1}{\sqrt{6}}\sum_{j\chi} \Phi_{0,j\chi}$ & identity\\
$\spG_{1'}^+$ & $-$ & $\frac{1}{\sqrt{6}}\sum_{j\chi} \chi \Phi_{0,j\chi}$ & pseudoscalar (\reffig{FigPinwheel})\\
$\spG_{12}^+$ & $d_{1+}$, $\bar{d}_{1+}$ & $\sum_{j\chi}v_j(d_1)\Phi_{0,j\chi}$, $\sum_{j\chi}v_j(\bar{d}_1)\Phi_{0,j\chi}$ &\\
$\spG_{12'}^+$ & $d_{1-}$, $\bar{d}_{1-}$ & $\sum_{j\chi}v_j(\bar{d}_1)\chi\Phi_{0,j\chi}$, $\sum_{j\chi}v_j(d_1)\chi\Phi_{0,j\chi}$ & see \reffig{FigSpinSpiral}\\
$\spX_1^+$ & $d_{j0}(-\frac{\pi}{2}\uv_j)$, $d_{j1}(-\frac{\pi}{2}\uv_j)$ & $\frac{1}{\sqrt{2}}(\Phi_{j,[j-1]+} + \Phi_{j,[j+1]-})$, $\frac{1}{\sqrt{2}}(\Phi_{j,[j+1]+} + \Phi_{j,[j-1]-})$\footnotemark[1]&\\
$\spX_2^+$ & $\bar{d}_{j0}(-\frac{\pi}{2}\uv_j)$, $\bar{d}_{j1}(-\frac{\pi}{2}\uv_j)$ & $\frac{1}{\sqrt{2}}(\Phi_{j,[j-1]+}-\Phi_{j,[j+1]-})$, $\frac{1}{\sqrt{2}}(\Phi_{j,[j+1]+}-\Phi_{j,[j-1]-})$\footnotemark[1]& see \reffig{FigBondOrdered}\\
$\spG_{15'}^-$ & $\Mv$ & & magnetization\footnotemark[2]\\
$\spX_3^-$ & $\helix_{j\chi}$ & $\Phi_{j,j\chi}$ & spin spiral (\reffig{FigSpinSpiral})
\end{tabular}
\end{ruledtabular}
\footnotetext[1]{In the subscript, $[j\pm 1]$ indicates arithmetic modulo $3$.}
\footnotetext[2]{The uniform magnetization does not correspond to any combination of these bilinears.}
\end{table*}

Linear combinations of the spins transform according to representations $\Gammam^{\kv p {-}}$, and there is in fact only one relevant representation for each of the stars $\star\kv_{\spG}$ and $\star\kv_{\spX}$. For the former, this is the pseudovector representation $\spG_{15'}$ (in the naming convention of Bouckaert;\cite{Cornwell} representation $T_{1g}$ of the point group $O_h$), and the corresponding observable is the uniform magnetization $\Mv$. As noted in \refsec{SecGaugeTheory}, the present analysis is restricted to situations where there is no net polarization, and so these order parameters vanish.

\begin{figure}
\putinscaledfigure{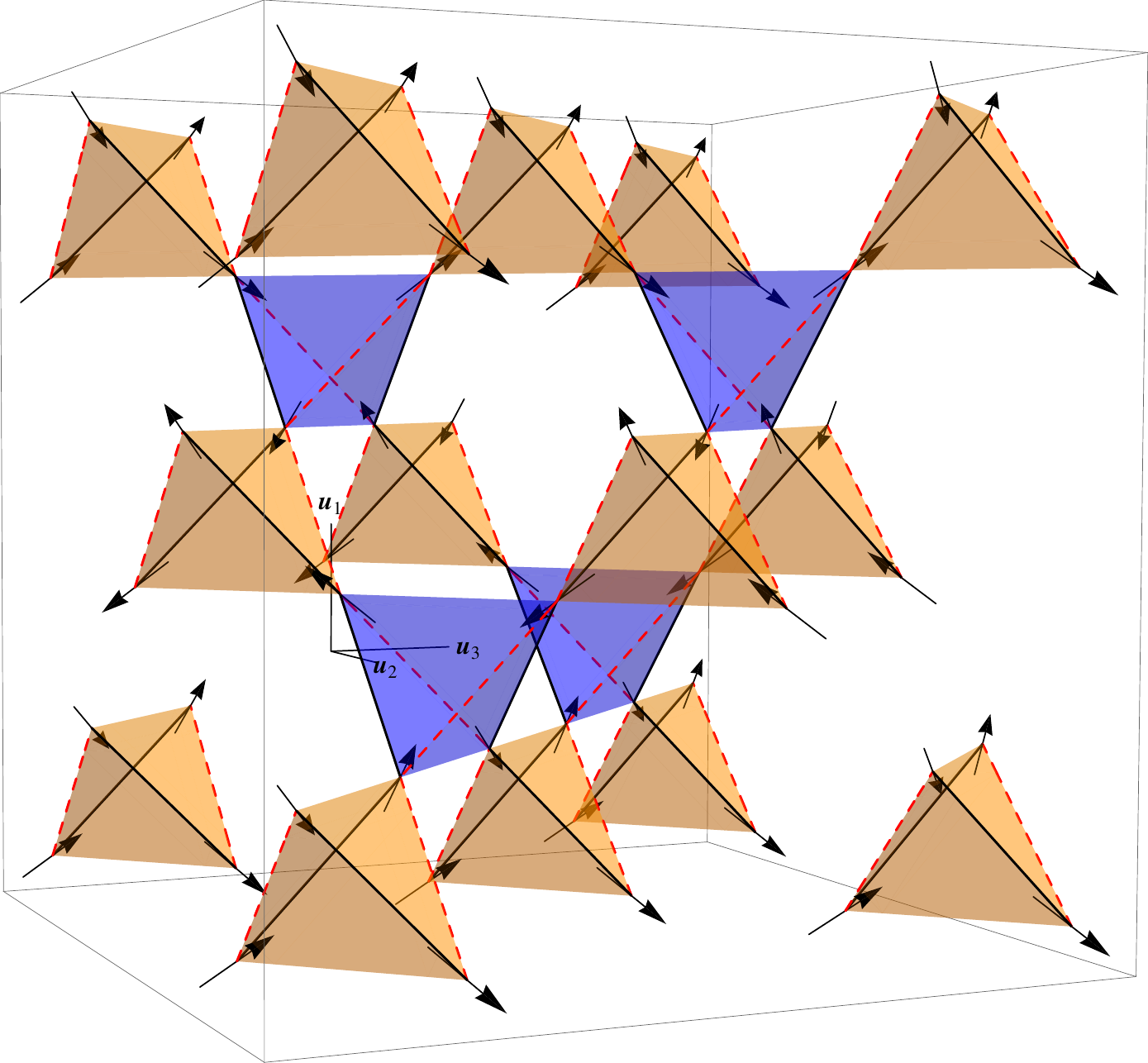}
\caption{A spin spiral maximizing $\helix_{1+}$ and an example of a perturbation that favors it. There are $12$ spiral-ordered states, with either sign for one of the $6$ order parameters $\helix_{j\chi}$ describing spirals of chirality $\chi = \pm$ along the cubic axis $\uv_j$. The observable labeled $\bar{d}_{1-}$ in \reftab{TabOrderParameters} is shown by solid (black) and dashed (red) lines, indicating respectively positive or negative signs for the correlations on these links. This spin spiral (or its time-reversed partner) also maximizes $\bar{d}_{1-}$: spins joined by solid and dashed bonds have, respectively, positive and negative inner product. An additional nearest-neighbor Heisenberg interaction with this symmetry (ferromagnetic on solid bonds, antiferromagnetic on dashed) therefore has energy minimized by the spin configuration shown.
\label{FigSpinSpiral}}
\end{figure}

At the star $\star\kv_{\spX}$, the irrep $\spX_3$ is associated with spin spirals, as illustrated in \reffig{FigSpinSpiral}. (Such states were found in the Monte Carlo studies of Melko et al.,\cite{Melko} who described them as consisting of antiferromagnetically stacked planes of tetrahedra; see \refsec{SecConclusions}.) This representation is $6$-dimensional, and the $6$ observables, denoted $\helix_{j\pm}$, are order parameters for right- and left-handed spirals along the $3$ cubic directions. The corresponding bilinears are given by simply $\Phi_{j,j\pm,j\pm}$. As noted in \refsec{SecRepsOthers}, improper transformations exchange momenta $\kappav_{j\pm}$ and hence the spiral chiralities.

The observables for irreps $\Gammam^{\kv p {+}}$ are somewhat more diverse. Consider first the $1$-dimensional representations at $\star\kv_{\spG}$, scalar $\spG_1$ and pseudoscalar $\spG_{1'}$. The former is fully symmetric and is given (for any star $\star\kappav$) by the symmetric combination of bilinears $\sum_\kappa \Phi_{0\kappa\kappa}$. As noted in \refsec{SecObservables}, an observable transforming as a pseudoscalar (invariant under proper transformations, but odd under spatial inversion) can be constructed using correlations of second-neighbor spins. The bilinear for the same irrep can be written $\sum_{j\chi} \chi \Phi_{0,j\chi,j\chi}$; it involves the difference between the bilinears for the wavevectors $\kappav_{j\chi}$ associated with opposite chirality $\chi$.

\begin{figure}
\putinscaledfigure{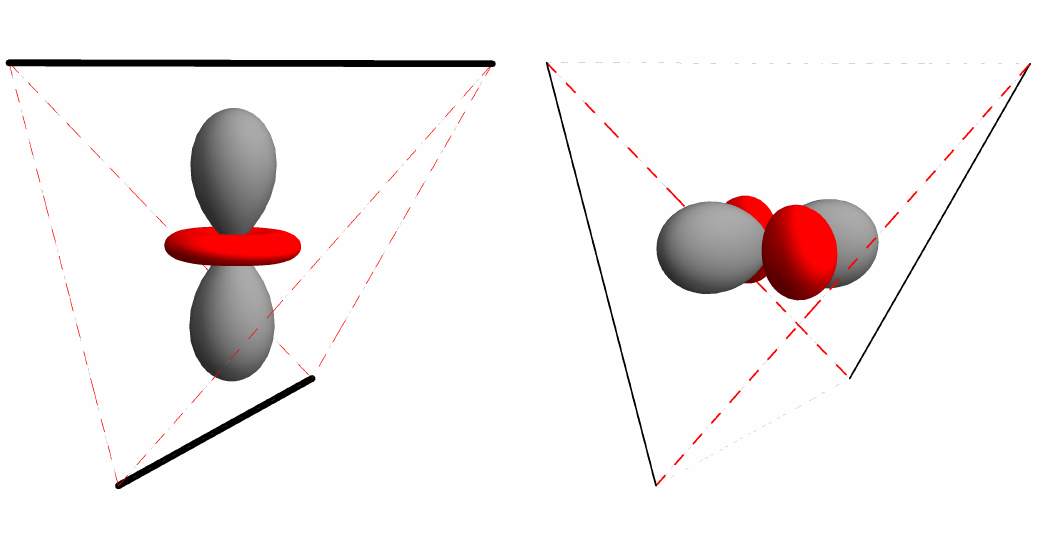}
\caption{Nematic order parameters labeled $d_j$ and $\bar{d}_j$ (where the cubic unit vector $\uv_j$ is vertical) for a single tetrahedron. These observables involve the illustrated linear combinations of nearest-neighbor spin correlations, where the line thickness indicates magnitude and dashed (red) lines have negative sign. The symmetries of these order parameters are those of the $E_g$ cubic harmonics, $d_{z^2}$ and $d_{x^2 - y^2}$ (where $z$ is vertical), which are plotted at the centers of the tetrahedra.
\label{FigdWave}}
\end{figure}

The remaining observables can be expressed in terms of the nematic variable $\Nv_i$, defined in \refsec{SecObservables}, measuring the nearest-neighbor correlations on tetrahedron $i$. For the star $\star\kv_{\spG}$, the relevant irreps are $\spG_{12}$ and $\spG_{12'}$ (representations $E_g$ and $E_u$ of $O_h$), which are both $2$-dimensional. The observables corresponding to both have spatial structures, illustrated in \reffig{FigdWave}, related to the $E_g$ cubic harmonics $d_{z^2}$ and $d_{x^2 - y^2}$. They will be denoted $d_{1+}$ and $\bar{d}_{1+}$ for the irrep $\spG_{12}$, and $d_{1-}$ and $\bar{d}_{1-}$ (see \reffig{FigSpinSpiral}) for $\spG_{12'}$; the subscript $\pm$ indicates that the sign for the correlations is the same or opposite on the two tetrahedron orientations. (As for the cubic harmonics, a different choice of ``quantization axis'' can be made by taking different linear combinations.) The observables for irreps $\spX_1^+$ and $\spX_2^+$ have wavevector $\kv = -\frac{\pi}{2}\uv_i$ and correlations restricted to tetrahedra of a single orientation. They have a similar $d$-wave structure, but with the symmetry axis along $\uv_i$.

As is evident from the case of the uniform magnetization, there is in general no guarantee that the observables and bilinears be in one-to-one correspondence. The intention is to include enough observables that there is at least one for every bilinear, and it is obviously permissible to ignore those observables that do not correspond to any bilinear. There might, however, be multiple sets of observables transforming in the same way, in which case any linear combination can be chosen, since all behave similarly. Alternatively, multiple combinations of bilinears may have the same transformation properties, and the observable will be given by some unknown combination of the given bilinears.

\section{Phase transitions}
\label{SecPhaseTransitions}

The nearest-neighbor model remains in a Coulomb phase at arbitrarily low temperatures, but perturbations can drive it into a Higgs phase. A standard Landau theory expressed directly in terms of the order parameters does not describe such a transition, but one can instead write down an effective action in terms of the coarse-grained fields $\phiv$ and $\alphav$. This contains, in principle, all terms compatible with symmetry, and can be expressed as a power series in successive powers of the fields and derivatives. The terms of most interest will in fact be those containing only the fields $\phiv$, and these are conveniently expressed in terms of the order parameters $\Phiv$.

This relies on the assumption that there is some nontrivial long-wavelength description, in terms of smooth variations on the minimal-energy configurations. The focus in the following will therefore be on cases where this is true, viz.\ for continuous transitions and also for weak first-order transitions, provided the correlation length becomes sufficiently large. Like a standard Landau theory, such an expansion can nonetheless be useful even when the transition is strongly first-order, where it can often at least exclude the possibility of a continuous transition.

Given the long-wavelength action, one can determine the possible transitions based on the behavior of the order parameters as the coupling constants are varied. Although certain ordered states and the perturbations that select them will be described, this should not be considered the main focus of the current work. Numerical simulations are better suited to address many such questions, which can be difficult to answer definitively using analytic arguments. Where the applied perturbations preserve the full space group, the approach will be to give an outline of the general types of transitions that are possible. Then specific examples will be given of symmetry-breaking perturbations for which the resulting ordered states and phase transitions can be determined.

\subsection{Long-wavelength action}
\label{SecLongWavelengthAction}

The aim is therefore to find a long-wavelength action (free-energy density) as an expansion in powers of the fields $\phiv$ and $\alphav$ and the derivative operator $\parv$, with each term symmetric under the space group $\spacegroup$ (or a subgroup, if a perturbation explicitly breaks symmetries).

The terms involving only the matter fields $\phiv$ are those that are gauge invariant, scalars under the space group, and time-reversal even. These can be found by considering all combinations of the bilinears $\Phi^{\kv p \pm}_a$ introduced in \refsec{SecBilinears}, and projecting into the identity representation. In the fully symmetric case, the only allowed term at first order in the bilinears is therefore given by the combination belonging to irrep $\spG_1$, labeled ``$+$'' in \reftab{TabOrderParameters}.

The higher-order terms are found by taking symmetric combinations of all products of the bilinears. A general procedure exists to decompose a product into its irreps (see, for example, \refcite{Lax}), but the projector for the identity representation can be constructed by simply summing the transformation matrices over the space group. The set of all second-order combinations of the bilinears can be written as $\Phiv\otimes\Phiv$, and the corresponding projector into the identity representation is given by an expression analogous to \refeq{EqProjD},
\beq{EqProjDeltaId}
\ProjDelta{}\sub{id}^{(2)} = \frac{1}{\lvert\spacegroup\rvert} \sum_{\sgQ\in\spacegroup}  \Deltam(\sgQ)\otimes\Deltam(\sgQ)\punc{,}
\eeq
where $\otimes$ denotes an outer product. Subsequent terms in the expansion can be found by taking correspondingly higher-order products.

For the fully symmetric case, the quartic terms in fact have a simple expression in terms of the bilinears $\Phi^{\Gamma}_a$ as
\beq{EqQuarticTerms}
\Action^{(4)} = \sum_{\Gamma} u_{\Gamma} \sum_a \lvert\Phi^{\Gamma}_a\rvert^2\punc{,}
\eeq
i.e., for each irrep $\Gamma$, a sum over the squares of the rows $a$. When perturbations reduce the symmetry, a similar result applies, but with the bilinears assigned to irreps of the appropriate subgroup of $\spacegroup$. This can be found by making use of a projector defined analogously to \refeq{EqProjDeltaId}, but with the sum over the subgroup.

The terms involving $\alphav$ and $\parv$ are strongly constrained by gauge symmetry, \refeq{EqGaugeContinuum}, and the lowest-order terms take the familiar form for a continuum gauge theory. They consist of a covariant-derivative term for each matter field (with ``minimal coupling'' to $\alphav$) and a pure gauge term, and can be viewed as deriving directly from the first two terms of the microscopic energy function $\Ham_{A,\theta}$ in \refeq{EqEA2}. The precise form of these terms can again be found by constructing the appropriate projection matrices, using the transformation properties of $\alphav$ and $\Dv$ specified in \refsec{SecContinuumFields}.

In the fully symmetric case, the long-wavelength action takes the form
\begin{multline}
\label{EqActionSymmetric}
\Action = \sum_{ij\chi} (1 + b\delta_{ij}) (D_i \phiv_{j\chi})^\dagger(D_i\phiv_{j\chi}) + K \lvert\parv \times \alphav\rvert^2\\
{}+ R\sum_{j\chi}\phiv_{j\chi}^\dagger \phiv_{j\chi}\ns + \Action^{(4)} + \cdots\punc{.}
\end{multline}
Terms involving more than two derivatives and matter fields beyond quartic order have been omitted, and are represented by the ellipsis.

This model has a Coulomb phase for sufficiently large $R$, and can exhibit various types of Higgs transitions when $R$ is reduced. The critical value $R\sub{c}$ and the nature of the ordered phase will depend on the coefficients of the terms appearing in $\Action^{(4)}$. Note that the action involves a total of $12$ complex fields $\varphi_{j\chi\ell}$ and the constant quadratic term in fact has full $\mathrm{SU}(12)$ symmetry. While this is reduced considerably by $S^{(4)}$, permutation symmetry remains between the $6$ wavevectors $\kappav$, and it is not possible in this case to eliminate any of the fields $\varphi_{j\chi\ell}$.

Higher-order terms than those displayed in \refeq{EqActionSymmetric} are sometimes necessary to determine the ordering completely. This occurs because the action including the quartic terms in \refeq{EqQuarticTerms} has continuous symmetry under rotations of the vector $\Phiv^\Gamma$ formed from the rows of a given irrep $\Gamma$ (of dimension larger than one), while the microscopic model has only discrete spatial symmetries. There can therefore be states with different orientations of $\Phiv^\Gamma$ within the continuous manifold that are not related by physical symmetries and that will be distinguished by higher-order terms in the action. (A concrete example will be described below in the case with reduced symmetry.) In the case of a continuous transition, the higher-order terms are likely to be ``dangerous irrelevant'' (strictly irrelevant at the transition but important away from it) and the continuous symmetry is emergent at the critical point.

Applying a perturbation coupling to one of the observables $F^\Gamma_a$ adds to $\Action$ a first-order coupling to the corresponding bilinear $\Phi^\Gamma_a$. For those perturbations that break no symmetries, the appropriate coupling is to the bilinear for the identity irrep $\spG_1^+$, and simply has the effect of changing the coefficient $R$. Symmetry-breaking perturbations, by contrast, lead to coefficients for the quadratic terms $\varphi_{j\chi\ell}^*\varphi_{j\chi\ell}\ns$ that depend on $j$, $\chi$, and (for those that break translation symmetry) $\ell$. They can therefore distinguish a subset of the fields $\varphi_{j\chi\ell}$ as relevant at the transition, while the others can be eliminated from (in principle, integrated out of) the long-wavelength action. Besides reducing the symmetry-implied restrictions on terms in the action, applied perturbations can therefore reduce the number of matter fields that must be included.

A particularly interesting case is where the symmetry is reduced to the extent that a single wavevector is relevant in the long-wavelength limit.\label{SecSingleWavevector} According to \reftab{TabOrderParameters}, the bilinear corresponding to the observable $\bar{d}_{1-}$ (illustrated in \reffig{FigSpinSpiral}) is
\begin{multline}
\label{Eqd1mbilinear}
\Phi^{\spG_{12'}^+}_2 = \frac{1}{\sqrt{3}}\left(\Phi_{0,1+} - \Phi_{0,1-}\right) \\{}+ \frac{1}{\sqrt{12}}\left(\Phi_{0,2-} + \Phi_{0,3-} - \Phi_{0,2+} - \Phi_{0,3+}\right)\punc{,}
\end{multline}
so it is sufficient to add a coupling to $\bar{d}_{1-}$ (with appropriate sign) to pick out the bilinears with $\kappav = \kappav_{1+}$. (It is of course possible for the same symmetries to be broken spontaneously instead, in which case the observable $\bar{d}_{1-}$ will become nonzero despite having no external coupling.)

With such a perturbation, the unbroken symmetries $\sgQ$ are those for which $\kappav_{\sgQ}(\kappav_{1+}) = \kappav_{1+}$, and can be determined by inspection of the links in \reffig{FigSpinSpiral}. They include rotations by $\pi$ about all three cubic axes $\uv_i$, and screw rotations by $\frac{\pi}{2}$ along the helices formed by links with a given sign. There are no remaining improper symmetries, since these exchange chiralities and relate wavevectors $\kappav_{j\chi}$ with opposite $\chi$.

The action is in this case given by
\begin{multline}
\label{EqActionReduced}
\Action\sub{R} = \sum_{i} (1 + b\delta_{i1}) (D_i \phiv_{1+})^\dagger(D_i\phiv_{1+}) + K \lvert\parv \times \alphav\rvert^2 \\
{}+ R\phiv_{1+}^\dagger \phiv_{1+}\ns + \Action\sub{R}^{(4)} + \cdots\punc{,}
\end{multline}
where other matter fields $\phiv_{j\chi}$ have been eliminated (renormalizing coefficients such as $R$), and coordinates along $\uv_1$ have been rescaled to make $K$ isotropic. The quartic terms can be written as
\beq{EqQuarticTermsReduced}
\Action\sub{R}^{(4)} = u_1 \Phi_1^2 + u_2 (\Phi_2^2 + \Phi_3^2)\punc{,}
\eeq
where $\Phi_{i} \equiv \Phi_{i,{1+},{1+}} = \phiv^\dagger_{1+}\sigmam^i\phiv_{1+}\ns$. Bilinears involving matter fields other than $\phiv_{1+}$ have been dropped, and the equality between $u_1$ and $u_2$ previously enforced by rotation symmetry no longer applies. Because of the reduced symmetry, certain observables such as spin spirals other than $\helix_{1+}$ can no longer be expressed in terms of relevant matter fields; these cannot therefore be order parameters for any transitions described by the action $\Action\sub{R}$. Referring to \reftab{TabOrderParameters}, one also finds that the observables $-$ and $d_{1+}$ couple to $\Phi_0 \equiv \Phi_{0,{1+},{1+}}$ and so are always nonzero (breaking no further symmetries).

The action $\Action\sub{R}$ is similar to that describing transitions of the cubic dimer model,\cite{CubicDimers,Charrier,Chen} having two relevant matter fields and quadratic terms that are symmetric under $\mathrm{SU}(2)$. Several symmetry-breaking scenarios in this model have been considered by Chen et al.,\cite{Chen} and use will be made of their results at several points in the following. (The same critical theory also describes the Mott insulator--superfluid transition for quantum bosons at half filling on a square lattice.\cite{Balents})

\subsection{Ordered states}
\label{SecOrderedStates}

In the models of \refeqand{EqActionSymmetric}{EqActionReduced}, a transition out of the Coulomb phase occurs when the matter field $\phiv$ condenses and the gauge field $\alphav$ acquires a finite correlation length by the Higgs mechanism. In both cases $\phiv$ has multiple components, so condensation also breaks certain spatial symmetries. Because of the gauge symmetry of $\phiv$, it is more convenient to describe these transitions in terms of the gauge-invariant bilinears $\Phiv$, which will therefore serve as order parameters. (It is possible to apply sufficient perturbations that there is a single relevant field $\varphi$ remaining, so there is no spontaneous symmetry breaking. An example will be treated briefly at the end of this section.)

In the fully symmetric case, the Higgs transition involves the matter fields condensing at some subset of the $6$ wavevectors $\kappav_{j\chi}$, with an orientation chosen for each two-component complex vector $\phiv_{j\chi}$ (up to gauge redundancy). This clearly admits the possibility of a direct transition where the matter field condenses into such a configuration. One could instead imagine a transition at which the Coulomb-phase correlations remain, but spatial symmetry is spontaneously broken, followed at a lower temperature by a Higgs transition of the remaining relevant matter fields. The latter will be identical to one with explicitly broken spatial symmetry, to be discussed below, and may be continuous or of first order.

\begin{figure}
\putinscaledfigure{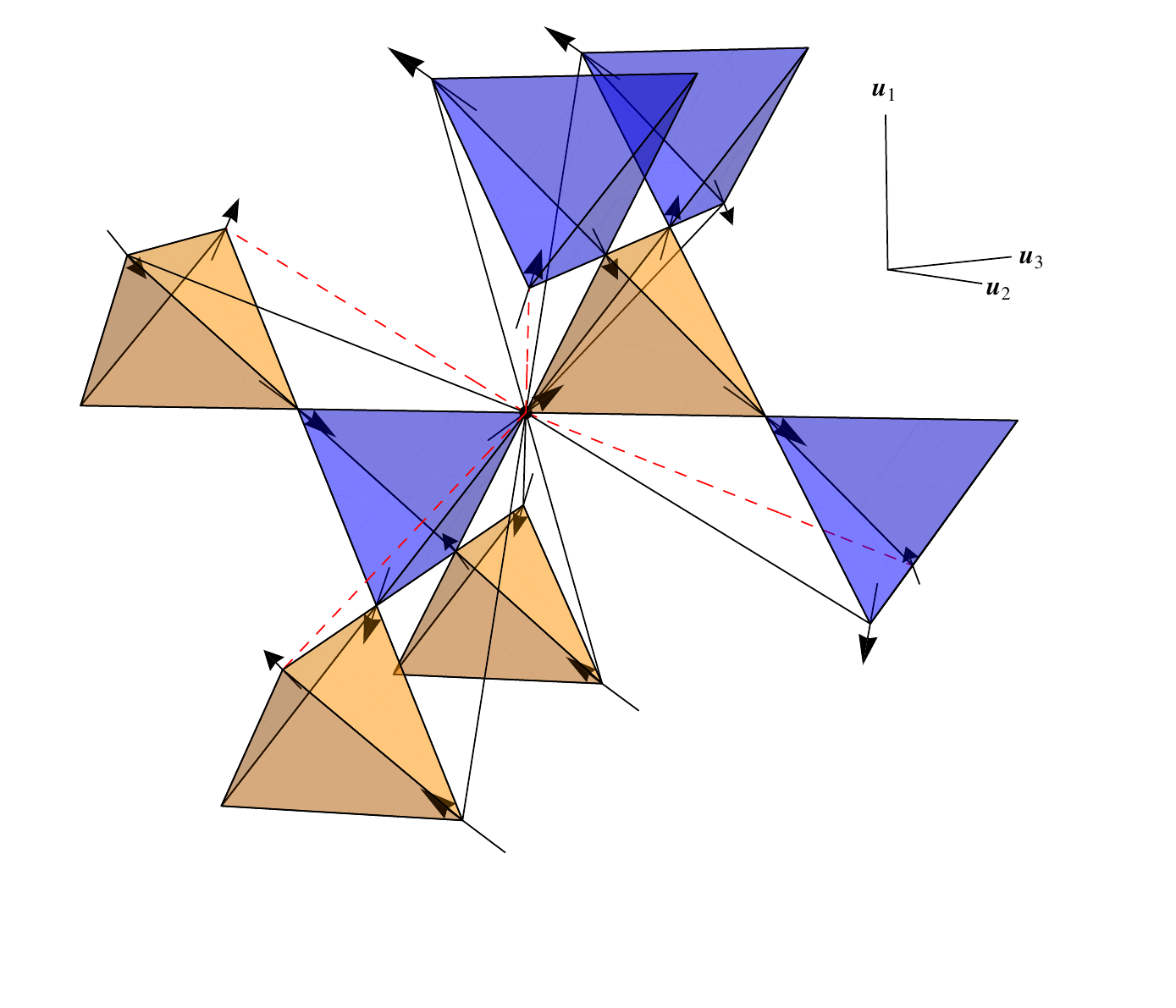}
\caption{Second-neighbor interactions between spins, as in \reffig{FigPinwheel}, favoring a spin spiral, as in \reffig{FigSpinSpiral}. Such a spiral, one of $12$ symmetry-related states, minimizes the energy if all second-neighbors have antiferromagnetic (AF) Heisenberg interactions (so no symmetries are explicitly broken). The $8$ solid (black) lines show second neighbors that satisfy this perturbation, while the $4$ shown dashed (and red) are unsatisfied. Considering all configurations of two adjacent tetrahedra (of $6$ pairs, at most $4$ are AF aligned) shows that one cannot satisfy more than ${2}/{3}$ of the second-neighbor links while obeying the ice rule. (It is worth noting a counterintuitive result of the geometry of spin ice: second neighbors that are both AF aligned with their common neighbor point towards tetrahedra of the same orientation and so are also AF aligned with each other.) Conversely, ferromagnetic interactions between second-neighbor spins disfavor the spiral configuration, and can instead lead to a bond-ordered state such as shown in \reffig{FigBondOrdered}.
\label{FigPinwheelSpiral}}
\end{figure}

An interaction between second-neighbor spins, as in \reffig{FigPinwheel}, provides a specific example of a perturbation that preserves all symmetries but can lead to an ordered state. It is demonstrated in \reffig{FigPinwheelSpiral} that antiferromagnetic interactions are minimized by arranging the spins into a spiral state (as in \reffig{FigSpinSpiral}), which is one of $12$ symmetry-related configurations. (The argument does not prove that these $12$ are the only minimal-energy configurations; additional interactions may in practice be necessary to stabilize the spiral state.) A spiral with axis along $\uv_j$ and chirality $\chi$ involves condensation of the matter field at wavevector $\kappav_{j\chi}$; the diagonal bilinear $\Phi_{j,j\chi,j\chi} = \phiv_{j\chi}^\dagger \sigmam^j \phiv_{j\chi}\ns$, corresponding to the observable $\helix_{j\chi}$, serves as an order parameter and can be positive or negative in the ordered phase.

In cases where a perturbation reduces the symmetry explicitly, the types of ordered states that are preferred will again depend on the details of the perturbation. As long as there is sufficient symmetry that more than one wavevector $\kappav$ is relevant to the ordering, one has a similar situation to the fully symmetric case.

The case where the single wavevector $\kappav_{1+}$ is selected has been considered above: the action is given in \refeq{EqActionReduced} and a suitable perturbation is illustrated in \reffig{FigSpinSpiral}. Also shown is a configuration of spins that minimizes the energy and may be selected by an ordering transition. The spins form a right-handed spiral described by the observable $\helix_{1+} \sim \Phi_1 = \phiv^\dagger_{1+}\sigmam^1\phiv_{1+}\ns$, using the notation of \refeq{EqQuarticTermsReduced}. There are two such states, related by time reversal (or the translation $\sgT_2$, for example), with opposite sign for $\Phi_1$.

One can therefore expect a transition in the presence of this perturbation into a Higgs phase with a nonzero value of the real order parameter $\Phi_1$. (One of the two linear combinations $\varphi_{1+,0} \pm \varphi_{1+,1}$ condenses.) This transition, at which an Ising symmetry is broken, resembles that in the ``2-GS'' model of Chen et al.,\cite{Chen} which was found to be of first order.

The form of the quartic terms in \refeq{EqQuarticTermsReduced} suggests the possibility of other ordering transitions, where the vector $\Phiv$ aligns elsewhere in the $3$-dimensional space. According to \refeq{EqPhiTheta}, the components $\Phi_2$ and $\Phi_3$ are even under time reversal, so states where only these are nonzero are paramagnetic; the appropriate order parameters can be found by reference to \reftab{TabOrderParameters}. Changing the relative magnitude of $u_1$ and $u_2$ tunes the system between these and the spin-ordered states with $\langle\Phi_1\rangle \neq 0$, and can be effected by further perturbations that do not reduce the symmetry but disfavor spin-ordering. An example is provided by the second-neighbor interactions illustrated in \reffig{FigPinwheel}; the argument of \reffig{FigPinwheelSpiral} implies that the spin spiral maximizes the energy when these are ferromagnetic.

\begin{figure}
\putinscaledfigure{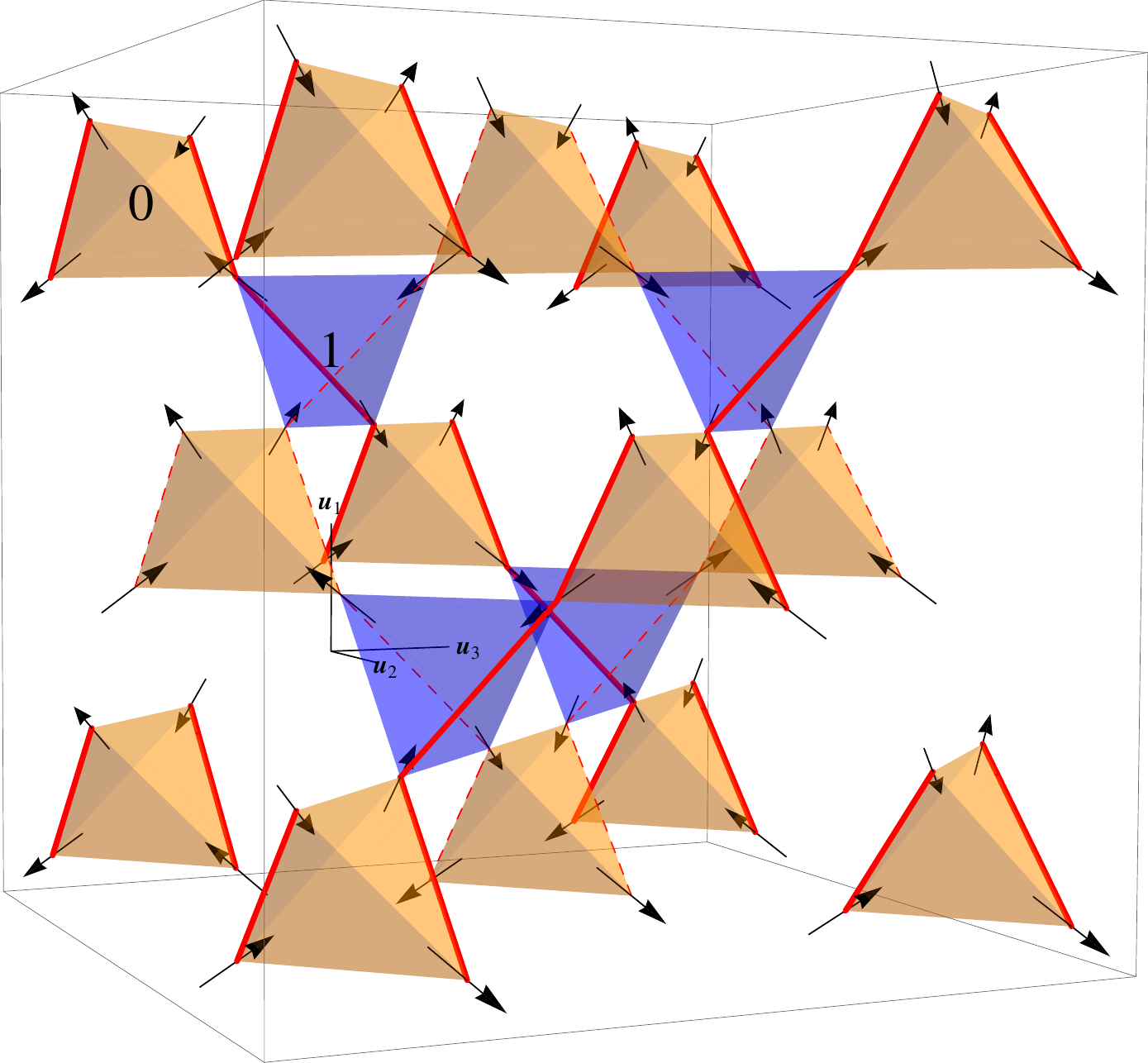}
\caption{Illustration of the bond-ordered phase, with order parameter $\Phi_{2,1+,1+}\sim\bar{d}_{20}(-\frac{\pi}{2}\uv_2)$, proposed in \refsec{SecOrderedStates} to occur in the presence of the perturbation shown in \reffig{FigSpinSpiral}. Only those links with additional antiferromagnetic (AF) interactions are shown (in red), and in this putative ordered state these links are divided into two types by their spin correlations. The thicker solid lines show links that are strongly AF aligned, i.e., pairs of spins with large negative expectation value for their inner product, while the thin dashed lines indicate weaker AF alignment. (If all the antiferromagnetic bonds were fully satisfied, the state would freeze into the spin spiral shown in \reffig{FigSpinSpiral}.) Symmetry under the translation $\sgT_2$ and the remaining $\frac{\pi}{2}$ screw rotations (along $\uv_1$) are broken, making the two tetrahedron orientations (labeled ``0'' and ``1'') inequivalent. The spins in the picture show an extreme example of such a configuration, where all favored antiferromagnetic bonds are satisfied and the remainder are unsatisfied; the actual ordered state has no spin ordering and a finite spin correlation length in all directions, including along the favored chains.
\label{FigBondOrdered}}
\end{figure}

An example of a state with $\langle\Phi_2\rangle \neq 0$ and $\langle\Phi_1\rangle = \langle\Phi_3\rangle = 0$ is illustrated in \reffig{FigBondOrdered}. According to \reftab{TabOrderParameters} (in particular, the rows for irreps $\spX_1^+$ and $\spX_2^+$), $\Phi_2 \equiv \Phi_{2,1+}$ can be associated with the observables $d_{20}(-\frac{\pi}{2}\uv_2)$ and $\bar{d}_{20}(-\frac{\pi}{2}\uv_2)$, which transform identically under the reduced symmetry group. In the ordered state shown, links with additional antiferromagnetic interactions are fully satisfied only in alternating $(010)$ planes, so the order parameter has wavevector $-\frac{\pi}{2}\uv_2$. For tetrahedra with $\sigma = 0$ (orange), $\bar{d}_2$ is nonzero and staggered on successive $(010)$ planes, so $\Phi_2 \sim \bar{d}_{20}(-\frac{\pi}{2}\uv_2) \neq 0$. By contrast, $\bar{d}_{2}$ vanishes on tetrahedra with $\sigma = 1$ (blue), so $\Phi_3 \sim \bar{d}_{21}(-\frac{\pi}{2}\uv_2) = 0$. (The single spin configuration shown is of course only a member of the ensemble; the state has no spin order.)

This phase resembles the intermediate paramagnets found by Chen et al.\cite{Chen}\ in the cubic dimer model. The latter occur when sufficient perturbations are applied that a Higgs transition occurs without spontaneously breaking any symmetries, followed by a lower-temperature transition into an ordered phase. The bond-ordered phase illustrated in \reffig{FigBondOrdered} does break spatial symmetries at the Higgs transition but preserves time-reversal symmetry and has no spin order. (It is also likely an intermediate phase: reducing the temperature further will remove the residual entropy leading to a spin-ordered state such as in \reffig{FigSpinSpiral}.)

The transition, by contrast, is apparently in the same class as the ``4-GS'' model of Chen et al.,\cite{Chen} described by a matter-field doublet $\phiv$ and easy-plane anisotropy for the bilinear $\Phiv$. Unlike the case where $\Phi_1$ becomes nonzero and a single component of the complex vector $\phiv_{1+}$ condenses, there is now a connected low-energy manifold, described by the phases of the two components of $\phiv_{1+}$. This led Chen et al.\cite{Chen}\ to conjecture that a continuous transition is possible, although a weakly first-order transition was found in their model.

Note that this example illustrates the observation made in \refsec{SecLongWavelengthAction} that higher-order terms are sometimes needed to specify the ordering. The quartic terms have a symmetry under continuous rotations mixing the $\Phi_2$ and $\Phi_3$ components of the vector $\Phiv$. (The quadratic term $\phiv_{1+}^\dagger\phiv_{1+}\ns = \Phi_0 = \lvert\Phiv\rvert$ is clearly also symmetric.) For $u_2 < u_1$, the free energy is minimized for $\Phiv$ lying anywhere in this plane, but this apparent degeneracy will be broken by higher-order terms such as $\Phi_2^2\Phi_3^2$, allowed by the discrete microscopic symmetries. (These higher-order terms also prevent the appearance of unphysical Goldstone modes due to the broken rotation symmetry.\cite{Senthil} All correlation functions are short-ranged in the ordered phase because the physical symmetries are discrete.)

A particularly interesting special case occurs when the quartic coefficients $u_1$ and $u_2$ are tuned to the same value. On the Higgs side of the transition, the order changes from one type to the other at this point, presumably in a strongly first-order ``spin flop''. At the multicritical point where this transition line meets the boundary of the Coulomb phase, the long-wavelength action given in \refeq{EqActionReduced} has an emergent $\mathrm{SU}(2)$ symmetry. There is then the possibility of a continuous transition in the ``noncompact $CP^1$'' class believed to describe ordering in the cubic dimer model.\cite{CubicDimers,Charrier,Chen,Papanikolaou,Charrier2}

A final example, where theoretical arguments strongly suggest a continuous transition, is when an applied perturbation favors a unique configuration such as the spin spiral shown in \reffig{FigSpinSpiral}. A suitable choice is clearly given by an applied magnetic field, coupling to the spins via a Zeeman term, with the same spatial configuration as the spiral. This perturbation breaks translation symmetry under $\sgT_2$ and $\sgT_3$ (as well as time reversal), so not only picks out a single wavevector, but also breaks the degeneracy in $\ell$. The critical theory can therefore be written in terms of a single scalar matter field $\varphi = \frac{1}{\sqrt{2}}(\varphi_{1+,0} + \varphi_{1+,1})$, and takes the standard form for a $\mathrm{U}(1)$ gauge theory. The Higgs transition breaks no spatial symmetries ($\langle\Phi_1\rangle > 0$, even in the Coulomb phase), and is in the usual inverted-XY universality class.\cite{Dasgupta}\label{SecTRBreaking}

\section{Conclusions}
\label{SecConclusions}

This work has analysed phase transitions out of the Coulomb phase in spin ice, and shown that they can be understood as Higgs transitions of an underlying gauge theory. A long-wavelength description results from finding the relevant modes of an emergent matter field, which transform according to representations of the magnetic symmetry group. The possible Higgs transitions are determined by forming order parameters from the matter fields and studying an effective action written in terms of these and the fluctuating gauge field. This stands in contrast to the standard paradigm of phase transitions, where the Ginzburg-Landau action is expressed directly in terms of the order parameters.\cite{Landau}

The phenomenology is particularly rich in the case of spin ice, as a consequence of symmetry under inversion of all spins (``time reversal''), which, as argued in \refsec{SecTimeReversal2}, leads to an extra (Kramers) degeneracy at the two highest-symmetry wavevectors. Since they cannot be dispersion minima of any hopping Hamiltonian (preserving time-reversal symmetry; see \refapp{AppHoppingModel}), a Higgs transition involving these wavevectors seems quite unlikely. The long-wavelength theory is instead expressed in terms of matter fields at $6$ wavevectors, leading to a variety of possible ordering transitions, both in the fully symmetric case and when applied perturbations reduce the number of relevant modes of the matter field. Because of this complexity, no attempt has been made here to describe all possible transitions or ordered states, and concrete results have been provided only in cases of particular interest. The analysis presented here can nonetheless be adapted to study other transitions in this system that may be discovered by experiments, numerics, or other theoretical approaches.

In fact, it is interesting to note that the Monte Carlo studies of Melko et al.\cite{Melko}\ found that the ground state in the presence of dipolar interactions is given by exactly the type of spin spiral shown in \reffig{FigSpinSpiral}. (As argued in \reffig{FigPinwheelSpiral}, this spiral is also favored by antiferromagnetic Heisenberg interactions between second neighbors.) The first-order transition observed in the numerical simulation is interpreted in the language of the present work as a Higgs transition of the matter field $\phiv_{j\chi}$ such that the diagonal bilinear $\Phi_{j,j\chi,j\chi}$ acquires a nonzero expectation value.

One straightforward prediction of the present work is a continuous transition in the inverted-XY universality class\cite{Dasgupta} into the spin spiral shown in \reffig{FigSpinSpiral}, in the presence of a helical magnetic field with the same spatial structure.\cite{EndNoteCQmapping} While this transition does not spontaneously break any symmetries, it is nonetheless an example of an exotic ``non-Landau'' transition, described by a critical theory of a single matter field coupled to an emergent $\mathrm{U}(1)$ gauge field. Producing a static magnetic field with the required structure on the atomic scale would of course be infeasible in an experiment, but trivial in a numerical simulation.

\begin{figure}
\putinscaledfigure{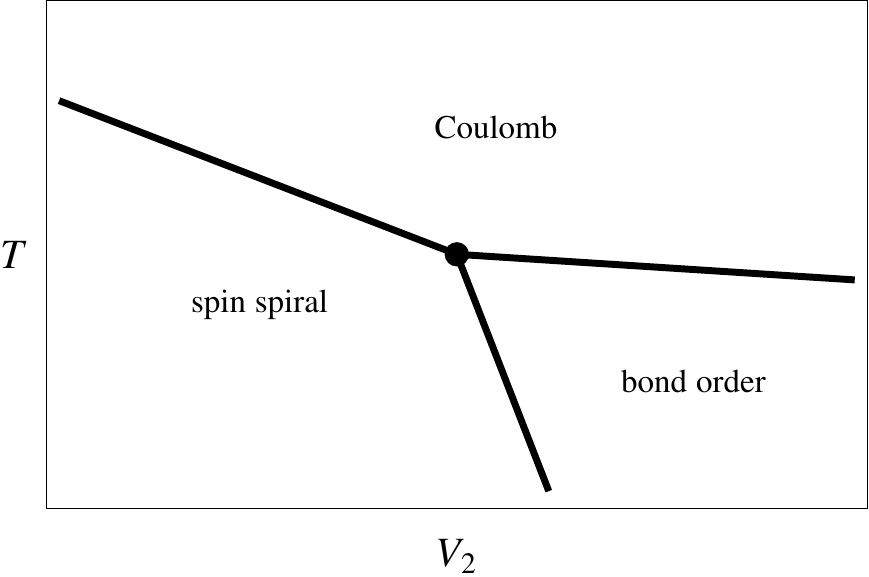}
\caption{Schematic phase diagram as a function of temperature $T$ and antiferromagnetic interaction $V_2$ between second-neighbor spins, in the presence of a (fixed) perturbation such as shown in \reffig{FigSpinSpiral}. The spin spiral, also shown in \reffig{FigSpinSpiral}, and the bond-ordered phase, \reffig{FigBondOrdered}, are ordered phases with distinct order parameters and are separated by a strongly first-order (``spin flop'') transition. The transitions into the Coulomb phase from the spin spiral and bond-ordered phase are described in \refsec{SecOrderedStates}. At the multicritical point where the three phases meet, there is an emergent $\mathrm{SU}(2)$ symmetry, allowing a continuous transition described by the noncompact $CP^1$ theory.
\label{FigSchematicPhaseDiagram}}
\end{figure}

An intriguing possibility that arises naturally from the analysis of \refsec{SecOrderedStates} is an emergent $\mathrm{SU}(2)$ symmetry at the point shown in \reffig{FigSchematicPhaseDiagram}, where the Coulomb phase meets two distinct ordered phases. The transitions into the two low-temperature phases, with spin-spiral and bond-based order parameters respectively, resemble those studied by Chen et al.\cite{Chen}\ in the cubic dimer model, which were shown to be of first order. At the point where the three phases meet, however, the long-wavelength theory given in \refeq{EqActionReduced} has an emergent $\mathrm{SU}(2)$ symmetry, allowing a continuous transition described by the noncompact $CP^1$ (NC$CP^1$) critical theory. If such a multicritical point is indeed found in numerics, its properties can be compared to those for other transitions\cite{Alet,Charrier,CubicDimers,Chen,Motrunich,Senthil,Balents} supposed to be described by NC$CP^1$. A clear signature of such a scenario would be an emergent $\mathrm{SO}(3)$ symmetry\cite{EndNoteSO5} in the correlations of the components of the vector order parameter $\Phiv$, as described in \refsec{SecOrderedStates}.

It must be noted that, even in these cases, the possibility of a first-order transition cannot be excluded on the basis of the analytic arguments presented here. Although the long-wavelength theory may exhibit a continuous transition (which remains controversial\cite{Kuklov2,Charrier2} for NC$CP^1$), this can be preempted by a first-order transition in a given microscopic model. In such cases, it may nonetheless be possible to tune the latent heat to zero using additional perturbations, as has been demonstrated explicitly in the cubic dimer model.\cite{Papanikolaou,Charrier2}

The analysis presented here excludes states with nonzero net polarization, which are favored whenever a uniform magnetic field is applied.\cite{Moessner111,Yoshida,Jaubert1,SpinIceCQ} (The ``squiggle'' state found to occur in quantum ice\cite{Shannon} also has nonzero, though unsaturated, polarization. Quantum effects can be included within the present approach by the addition of interactions favoring ``flippable'' configurations.) To remove this restriction, one must include the possibility of nonzero uniform flux in \refeq{EqBfromA}, which has the effect of modifying the background gauge field $a$. Results for this case will be presented elsewhere.

This work has also treated only the case where the ice rule is uniformly obeyed, and \refeq{EqDivB} can be viewed as a constraint. This is equivalent to forbidding magnetic monopoles,\cite{Castelnovo} and requires temperature much lower than the scale of the nearest-neighbor interactions. Deviations from this limit can be treated at a simple level by analogy with finite-size scaling: the analysis presented here is strictly applicable only in the thermodynamic limit with zero monopole density, but has relevance even when the system size and mean monopole separation are finite but large.

\acknowledgments

I am grateful to Leon Balents, John Chalker, Gang Chen, Sankar Das Sarma, Michael Levin, and Krishnendu Sengupta for stimulating discussions and helpful comments. This work is supported by JQI-NSF-PFC and DARPA QuEST.

\appendix

\section{Calculation of MSG commutation relations}
\label{AppMSGCommutation}

The commutation relation, \refeq{EqTcommutation}, of the FCC translation operators $\msgT_1$ and $\msgT_2$ was calculated in \refsec{SecMSG} by application of the relation \refeq{EqDefinepi2}. Commutation relations for other operators $\msgQ$ can be found by the same approach, with the subtlety that these depend on the arbitrary choices of global phase made for the functions $\pi_{\msgQ}$.

For example, one has $\sgR_1 \sgT_1 = \sgT_2 \sgR_1$ (where $\sgR_1$ is a proper rotation by $2\pi/3$ about the diamond link $\deltav_1$, with corresponding MSG operator $\msgR_1$), so the relative phase of $\msgT_1$ and $\msgT_2$ can be chosen such that $\msgR_1\msgT_1 = \msgT_2\msgR_1$. In other words, one fixes
\beq{EqR1T1T2R1}
\pi_{\msgR_1}(\sgT_1 j)\pi_{\msgT_1}(j) = \pi_{\msgT_2}(\sgR_1 j)\pi_{\msgR_1}(j)\punc{,}
\eeq
by appropriate choice of the relative phase of $\pi_{\msgT_1}$ and $\pi_{\msgT_2}$. Using \refeq{EqDefinepi2}, this gives
\beq{EqpiT1T2}
\pi_{\msgT_1}\ns(j)\pi_{\msgT_2}^*(\sgR_1 j) = \omega^*(\link j {\sgT_1 j}) \omega(\link{\sgR_1 j}{\sgR_1 \sgT_1 j})\punc{,}
\eeq
which determines other commutation relations involving $\msgT_1$ and $\msgT_2$. One can similarly fix the phase of $\msgT_3$ relative to $\msgT_1$ and $\msgT_2$ by requiring $\msgR_1 \msgT_3 = \msgT_1 \msgR_1$.

While these relations fix the relative phase of the translation operators $\msgT_i$, their absolute phases can be fixed using commutation relations such as $\sgT_1 \sgR_2 = \sgR_2 \sgT_2^{-1} \sgT_1$. The corresponding expression in terms of the MSG operators depends on the absolute phase of $\msgT_2$, as confirmed by considering
\begin{widetext}
\begin{align}
\pi_{\msgR_2^{-1} \msgT_1^{-1} \msgR_2\ns \msgT_2^{-1} \msgT_1\ns}(j) &= \pi^*_{\msgR_2}(j)\pi^*_{\msgT_1}(\sgR_2 j) \pi\ns_{\msgR_2}(\sgT_2^{-1}\sgT_1 j) \pi^*_{\msgT_2}(\sgT_2^{-1}\sgT_1 j) \pi\ns_{\msgT_1}(j)\\
&=\pi^*_{\msgT_2}(\sgT_2^{-1}\sgT_1 j) \omega(\sgT_1 j \linkarrow \sgT_1 \sgR_2 j \linkarrow \sgR_2 j \linkarrow j \linkarrow \sgT_2^{-1}\sgT_1\ns j)\punc{,}
\label{EqpiR2b}
\end{align}
where \refeq{EqDefinepi2} has been used to eliminate $\pi_{\msgR_2}$ and $\pi_{\msgT_1}$. Since $\msgR_2^{-1} \msgT_1^{-1} \msgR_2\ns \msgT_2^{-1} \msgT_1\ns = \pm\msgE$, this expression cannot depend on $j$, and so one can simplify the chain appearing in \refeq{EqpiR2b} by making a suitable choice of $j$. It is particularly convenient to set $j = \sgT_1^{-1}\sgT_2\ns0$, where $0$, the origin of the coordinate system, obeys $\sgR_2 0 = 0$. In this case, one has simply
\beq{EqpiR2c}
\pi_{\msgR_2^{-1} \msgT_1^{-1} \msgR_2\ns \msgT_2^{-1} \msgT_1\ns}(\sgT_1^{-1}\sgT_2\ns0) = \pi_{\msgT_2}^*(0) \omega(\link{\sgT_2 0}0)\punc{.}
\eeq
Since the absolute global phase of $\pi_{\msgT_2}$ is as yet unspecified, one can choose $\pi_{\msgT_2}(0)\omega(\link0{\sgT_20}) = 1$, giving $\pi_{\msgR_2^{-1} \msgT_1^{-1} \msgR_2\ns \msgT_2^{-1} \msgT_1\ns} = 1$, or equivalently $\msgT_1 \msgR_2 = \msgR_2 \msgT_2^{-1} \msgT_1$.

Note that the quantity
\beq{EqDefinephi}
\phi_{\msgT_i}(j) = \pi_{\msgT_i}(j) \omega(\link{j}{\sgT_i j})
\eeq
is invariant under changes of the gauge of the background field $a$, with $\pi_{\msgQ}$ transforming to maintain \refeq{EqDefinepi}. The choice made after \refeq{EqpiR2c} gives $\phi_{\msgT_2}(0) = 1$; the same is true for $\msgT_1$, according to \refeq{EqpiT1T2}, and for $\msgT_3$ similarly. Furthermore, \refeq{EqDefinepi2} implies
\beq{EqphiTi}
\phi_{\msgT_i}(j) = \phi_{\msgT_i}(0) \omega(0 \linkarrow \cdots \linkarrow j \linkarrow \sgT_i j \linkarrow \cdots \linkarrow \sgT_i 0 \linkarrow 0)\punc{;}
\eeq
the chain appearing on the right-hand side is a closed loop (`parallelogram') consisting of two pairs of parallel segments. Counting the number of enclosed hexagons gives
\beq{EqphiTij}
\phi_{\msgT_i}(j) = (-1)^{\frac{1}{2}(\uv_i \cdot \rv_j - \sigma_j)}\punc{,}
\eeq
where $\rv_j$ is the position vector of site $j$ and $\sigma_j \in \{0,1\}$ is its sublattice. Note that this quantity is real for all sites $j$ of the diamond lattice.

The commutation relation of a primitive translation $\msgT_i$ and any other MSG operator $\msgQ$ can be found by first considering the corresponding expression for the space-group operators $\sgT_i$ and $\sgQ$. It is simple to show that the combination $\sgQ \sgT_i \sgQ^{-1}$ is a pure translation $\sgT_{\MQ \ev_i}$, by displacement $\MQ \ev_i$. One is then led to consider the combination $\msgT_{\MQ \ev_i}^{-1}\msgQ\msgT_i\msgQ^{-1}$, and to compute
\beq{EqpiTQTQ}
\pi_{\msgT_{\MQ \ev_i}^{-1}\msgQ\msgT_i\msgQ^{-1}}(j) = \pi_{\msgT_{\MQ \ev_i}^{-1}}^*(j) \pi\ns_{\msgT_i}(\sgQ^{-1}) \omega^*(\link{\sgT_i\sgQ^{-1}j}{\sgQ^{-1}j}) \omega(\link{\sgT_{\MQ \ev_i}^{-1} j}{j})\punc{,}
\eeq
\end{widetext}
where \refeq{EqDefinepi2} has been used to eliminate $\pi_{\msgQ}$. Both chains in this expression have total displacement equal to a minimal FCC vector and can consistently be taken to follow the conventional path, as in \refeq{EqscTexpand}. One therefore finds
\beq{EqpiTQTQ2}
\pi_{\msgT_{\MQ \ev_i}^{-1}\msgQ\msgT_i\msgQ^{-1}}(j) = \phi_{\msgT_{\MQ \ev_i}}^*(j) \phi_{\msgT_i}(\sgQ^{-1}j)\punc{,}
\eeq
which is valid for all sites $j$, but, since $\sgT_{\MQ \ev_i}^{-1}\sgQ \sgT_i \sgQ^{-1} = \sgE$, must in fact be independent of $j$. Taking $j = 0$ gives $\rv_{\sgQ^{-1}0} = -\MQ^{-1}\VQ$, and one can choose the sign of $\msgT_{\MQ \ev_i}$ to make $\phi_{\msgT_{\MQ \ev_i}}(0) = 1$, so
\beq{EqpiTQTQ3}
\msgQ\msgT_i = (-1)^{\frac{1}{2}(\VQ\cdot \MQ\uv_i + \sigmaQ)}\msgT_{\MQ \ev_i}\msgQ\punc{,}
\eeq
using \refeq{EqphiTij}.

The MSG cubic translation $\msgK_i$ defined in \refeq{EqDefineKi} can alternatively be expressed as
\beq{EqDefineKib}
\msgK_i = -\msgT_{[i-1]}\ns \msgT_{i}^{-1} \msgT_{[i+1]}\ns\punc{,}
\eeq
where $[i\pm 1]$ appearing in the subscript indicates arithmetic modulo $3$. Analogous to $\phi_{\msgT_i}$, defined in \refeq{EqDefinephi}, the quantity
\beq{EqDefinephiK}
\phi_{\msgK_i}(j) = \pi_{\msgK_i}(j) \omega(j \linkarrow \sgT_{[i+1]}j \linkarrow \sgT_i^{-1}\sgT_{[i+1]}\ns j \linkarrow \sgK_i j)
\eeq
is gauge invariant. The chain $\omega(j \linkarrow \cdots \linkarrow \sgK_i j)$, consisting of minimal FCC translations interpreted according to the convention used in \refeq{EqscTexpand}, is chosen to comprise a right-handed helix as in \reffig{FigGaugeFieldDiagram}. Expressing $\pi_{\msgK_i}$ in terms of $\pi_{\msgT_i}$ with \refeqand{EqpiMultiplication}{EqDefineKib} and then using \refeq{EqphiTi}, this gives
\begin{align}
\label{EqphiK2a}
\phi_{\msgK_i}(j) &= -\phi\ns_{\msgT_{[i-1]}}(\sgT_{[i+1]}\ns\sgT_i^{-1}j) \phi^*_{\msgT_i}(\sgT\ns_{[i+1]}\sgT_i^{-1}j) \phi\ns_{\msgT_{[i+1]}}(j)\\
&= 1\punc{,}
\label{EqphiK2b}
\end{align}
the result expressed as \refeq{EqphiK} in \refsec{SecCubicTranslations}.

\section{Microscopic hopping model}
\label{AppHoppingModel}

A significant part of the present work involves finding representations of the magnetic symmetry group (MSG), to which the important modes of the matter field $\psi$ belong. Certain results of this analysis can be confirmed by studying a solvable model with the same spatial symmetries and finding the degeneracies and transformation properties of its modes. An obvious choice is to use the analogy, noted in \refsec{SecQuantumAnalogy}, with a quantum model of noninteracting bosons on the diamond lattice in the presence of a static gauge field $a_\lambda$. In this appendix, the spectrum of this (three-dimensional) Hofstadter problem\cite{Hofstadter} will be presented.

Starting from the model of \refeq{EqEA2}, softening the unit-modulus constraint on $\psi = \ee^{\ii \theta}$ and dropping the fluctuating gauge field $A_\lambda$ leaves a model with the appropriate symmetries and hopping terms only between nearest-neighbor sites. It is nonetheless worthwhile to include further-neighbor hopping, and to study a generic model with the same symmetries. As will be noted below, the nearest-neighbor problem has additional ``accidental'' degeneracies that cannot be expected to survive in the presence of the fluctuating gauge field.

\subsection{Hopping phases}

The hopping model is written in terms of complex fields $\psi_i$ and has Hamiltonian
\beq{EqHoppingHam}
\Ham\sub{hopping} = -\sum_{ij} t_{ij} \psi_j^* \psi_i\ns\punc{,}
\eeq
where the (dual diamond) sites $i$ and $j$ need not be neighbors and every pair is counted twice. For any space-group transformation $\sgQ$, the hopping amplitudes $t_{ij}$ transform under the corresponding MSG operator $\msgQ$, and so obey
\beq{EqHoppingMSG}
\begin{aligned}
t_{\sgQ i,\sgQ j} &= t_{ij} \pi_{\msgQ}^*(i) \pi_{\msgQ}\ns(j)\\
&= t_{ij} \omega(i \linkarrow \cdots \linkarrow j) \omega^*(\sgQ i \linkarrow \cdots \linkarrow \sgQ j)\punc{,}
\end{aligned}
\eeq
where \refeq{EqDefinepi2} has been used. In order for the energies to be real, one requires $t_{ji}\ns = t_{ij}^*$.

For nearest-neighbor links, hopping phases obeying \refeq{EqHoppingMSG} can be extracted directly from the first term of \refeq{EqEA2}: for these links, take $t_{ij} = \lvert t_{ij}\rvert \ee^{-2\pi \ii a_{\link ij}} = \lvert t_{ij}\rvert \omega^*(\link ij)$. A consistent set of phases for the further-neighbor hopping is given by
\beq{EqHoppingPhases}
t_{ij} = \frac{t_{ij}\zero}{\lvert\mathfrak{C}_{ij}\rvert}\sum_{C \in \mathfrak{C}_{ij}} \omega^*(C)\punc{,}
\eeq
where $t_{ij}\zero$ is real and $\mathfrak{C}_{ij}$ is a complete set of symmetry-related paths from $i$ to $j$. In other words, start by choosing an arbitrary path $C_1 = i \linkarrow \cdots \linkarrow j$, and sum over all space-group operators $\sgQ$ such that $\sgQ C_1$ also runs from $i$ to $j$ (i.e., $\sgQ i = i$ and $\sgQ j = j$). It is then straightforward to show that \refeq{EqHoppingMSG} is satisfied, provided that the magnitudes $t_{ij}\zero$ are symmetric under the space group.

For certain pairs of sites, the sum in \refeq{EqHoppingPhases} vanishes and the hopping $t_{ij}$ must vanish. This occurs, for instance, when the two sites are at opposite ends of a hexagon of the diamond lattice. The set $\mathfrak{C}_{ij}$ can be taken as consisting of the two shortest paths around the hexagon, but, as noted in \refsec{SecMSG}, these have $\omega$ differing by a minus sign, so the sum vanishes.

To treat the case where perturbations reduce the symmetry of the problem, one can modify the hopping amplitudes to obey \refeq{EqHoppingMSG} for the appropriate subgroup. Time reversal $\Theta$ acts according to \refeq{EqTRpsi}, and is a symmetry of the hopping Hamiltonian only if $\pi_\Theta$ can be found so that
\beq{EqHoppingTR}
t_{ij} \pi_{\Theta}\ns(i) \pi_\Theta^*(j) = t_{ij}^*
\eeq
for all sites $i$ and $j$. Taking the product of both sides for the links $\link i j$ around a hexagonal plaquette $\pi$, this implies that $\prod_{(\link i j) \in \pi} t_{ij}$ must be real if $\Theta$ is unbroken. This is indeed the case for the phases $\omega(\link i j) = \ee^{2\pi \ii a_{\link i j}}$, for which the product is always $-1$. Perturbations that break $\Theta$ (such as an applied magnetic field) are equivalent in the hopping problem to a nontrivial magnetic flux (not zero or half a flux quantum) through the hexagonal plaquette.

\subsection{Spectrum}

Once the hopping phases are determined, the Hamiltonian $\Ham\sub{hopping}$ can be expressed in the basis of eigenstates of $\msgK_i$. The structure of the energy eigenstates within the cubic unit cell can be found from the effective $\kappav$-dependent hopping
\beq{EqKappaHopping}
\tilde{t}_{ij}(\kappav) = \sum_{\sgK} t_{\sgK i,j} \pi_{\msgK}(i) \ee^{i \Rv_{\sgK}\cdot \kappav}\punc{,}
\eeq
where $i$ and $j$ are both sites within the cubic unit cell of the diamond lattice. The sum is over all cubic translations $\sgK$ by displacement $\Rv_{\sgK}$.

\begin{figure}
\putinscaledfigure{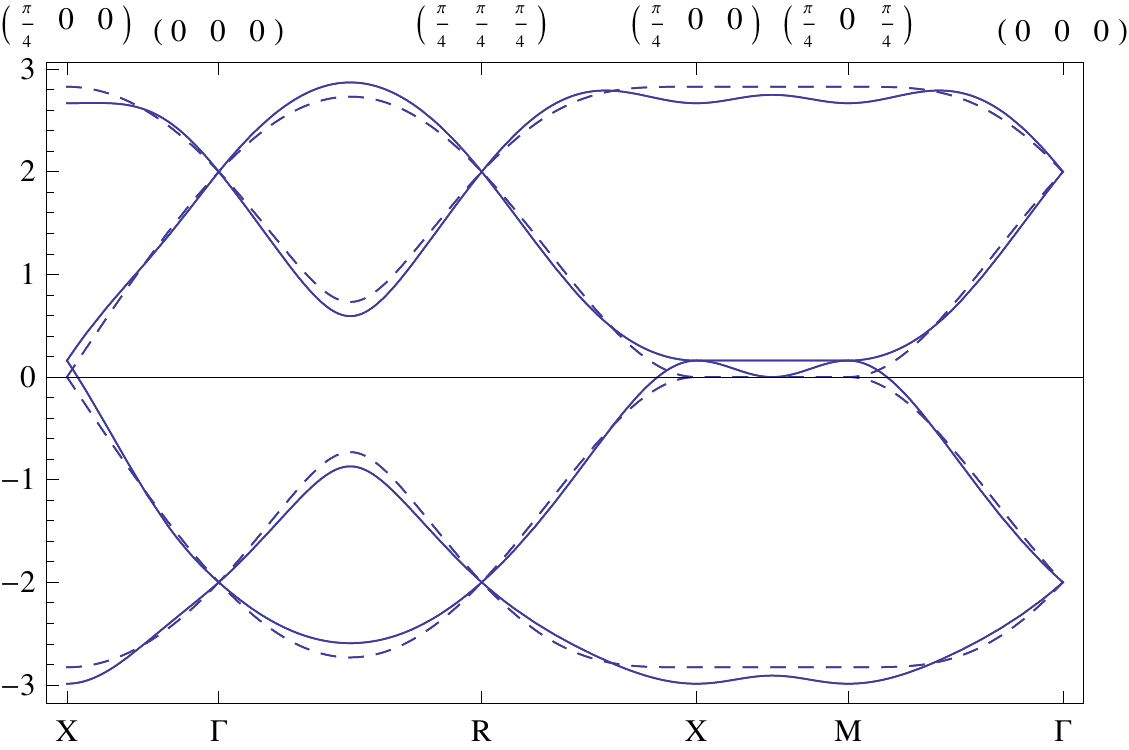}
\caption{Mode energy (in units of the nearest-neighbor hopping amplitude) in a hopping model with the full symmetry of the diamond lattice, as a function of wavevector $\kappav$ within the cubic reduced Brillouin zone $\BZR$. There are $8$ sites within the cubic unit cell, and hence $8$ bands, but there is a twofold degeneracy due to the anticommutation of the FCC translations $\msgT_i$. As discussed in \refsec{SecTimeReversal2}, there are additional degeneracies at the points $\spG$ and $\mathrm{R}$, which are each surrounded by (three-dimensional) Dirac cones, visible here as crossings. The dashed line shows the dispersion with hopping only between nearest-neighbor sites, which has flat bands along the lines $\mathrm{Z}$ joining points $\mathrm{X}$ and $\mathrm{M}$. (Throughout this appendix, all labels for symmetry points are referred to $\BZR$.) For nonzero dispersion along this line, one requires hopping between at least sixth neighbors; the solid line has a weak additional hopping between sites separated by $2\ev_1 + \ev_2 - \ev_3$ and equivalent vectors.\label{FigHoppingDispersion}}
\end{figure}

The dispersion for the fully-symmetric case, shown in \reffig{FigHoppingDispersion}, exhibits various features anticipated in \refsec{SecRepresentations}. Every mode is twofold degenerate, labeled by $\ell = 0, 1$, and the energies at wavevectors $\kappav$ and $\kappav_{\sgQ}(\kappav)$ are equal for all $\sgQ$. In particular, inversion symmetry implies degeneracy between the states at $\kappav$ and $\reducedR{\kappav - \frac{\pi}{4}\deltav_0}$, such as $\spG$ and $\mathrm{R}$. As noted in \refsec{SecTimeReversal2}, there are Dirac cones in the spectrum at the latter two points. The global minima of the spectrum are at the points $\spX$ and $\mathrm{M}$, which, as noted in \refsec{SecRepsOthers}, form a star containing $6$ wavevectors. (With only nearest-neighbor hopping, the energy is equal along the line $\mathrm{Z}$ between these points, but this accidental degeneracy is lifted by further-neighbor hopping.)

The symmetry-breaking perturbation introduced in \refsec{SecSingleWavevector} allows a single wavevector $\kappav_{1+}$ to be distinguished from the rest. This can be confirmed by calculating the spectrum with an additional hopping term that breaks the same symmetries as the perturbation shown in \reffig{FigSpinSpiral}. With the appropriate choice of sign for this hopping, there is indeed a unique minimum energy at the wavevector $\kappav_{1+}$.

A perturbation that selects a single spiral state must break time-reversal symmetry as well as the FCC translations $\sgT_i$, and corresponds to additional hopping phases in this model. The applied magnetic field favoring the spin order shown in \reffig{FigSpinSpiral} can be produced with a vector potential having the same spiral structure and the same cubic unit cell as the ordered state itself. The resulting spectrum is similar to the previous case, with a unique minimum at $\kappav_{1+}$, but with the twofold degeneracy also broken.


\begin{thebibliography}{99}
\bibitem{EndNoteCharges} One can instead justify the introduction of the charges using a dual description in terms of an effective electric field. Because the magnetic field is discrete, the electric field is compact (defined only modulo $2\pi$) and hence allows monopoles. These electric monopoles are exactly the matter fields described by $\theta$.
\bibitem{EndNotePSG} This is sometimes known as the ``projective symmetry group'',\cite{Wen} since it provides a projective representation of the space group.
\bibitem{EndNoteDoubleGroup} The multiplication rules for the MSG operators are equal to those of the space-group operators up to a sign, so the group consisting of $\pm \msgQ$ is by definition a ``double group''.\cite{Cornwell} (Such groups appear in the context of fermions, where certain rotations change the sign of the spinor wavefunction.) The relevant representations are those where the multiplication rules have the same sign as those of the MSG operators, the so-called ``extra'' representations of the double group. The analogy with the fermionic case is not complete, however, since the multiplication table is different here from the fermion case (where, for example, all translations commute), and so tabulated extra representations are of no use.
\bibitem{EndNoteCQmapping} This transition can also be understood using the classical-to-quantum mapping of \refciteand{Jaubert1}{SpinIceCQ}. The nonuniform magnetic field maps to a potential energy distinguishing the two sublattices of the resulting quantum model, and the spin spiral becomes a Mott insulator. The quantum phase transition from the superfluid (corresponding to the Coulomb phase) is in the standard $2+1$D XY universality class,\cite{Sachdev} in agreement with the present conclusions. (The transition is ``inverted'' because the superfluid, the ordered phase of the quantum model, maps to the high-temperature Coulomb phase of the classical model.)
\bibitem{EndNoteSO5} This resembles the $\mathrm{SO}(5)$ scenario proposed for the cuprates,\cite{Demler} where two apparently unrelated order parameters form a vector with an approximate symmetry. In the present case, there is an exact emergent $\mathrm{SU}(2)$ symmetry precisely at the multicritical point, provided that higher-order terms are (as seems likely) irrelevant.
\end{thebibliography}
\end{document}